\documentclass[%
 reprint,
 amsmath,amssymb,
 aps, pra, floatfix
]{revtex4-2}

\usepackage{graphicx}% Include figure files
\usepackage{dcolumn}% Align table columns on decimal point
\usepackage{bm}% bold math
\usepackage{xcolor}
\usepackage{hyperref}% add hypertext capabilities

\usepackage{tikz}
\usepackage{lipsum}
\usepackage{mathtools}  
\usepackage{xfrac}  
\usetikzlibrary{decorations.pathreplacing,calc,positioning,intersections}
\usetikzlibrary{patterns}
\usetikzlibrary{arrows,shapes,positioning,shadows,trees}
\usetikzlibrary{3d} % <--- use of canvas and rotation
\usetikzlibrary {shapes.arrows}
\usetikzlibrary{arrows.meta}
\usepgflibrary {shadings}
\definecolor{bleudefrance}{rgb}{0.19, 0.55, 0.91}
\definecolor{cadetgrey}{rgb}{0.57, 0.64, 0.69}
 
\usepackage{amsmath}
\usepackage{tikz}
\usetikzlibrary{arrows}
\definecolor{darkred}{rgb}{0.58, 0.075, 0}
\definecolor{lightred}{rgb}{1, 0.129, 0}

\begin{document}

\preprint{APS/123-QED}

\title{Target Field Design of Surface Permanent Magnets}

\author{Peter J. Hobson}
\author{Chris Morley}
\email{Christopher.Morley2@nottingham.ac.uk}
\author{Alister Davis}
\author{Thomas Smith}
\author{Mark Fromhold}
\email{Mark.Fromhold@nottingham.ac.uk}
\affiliation{School of Physics \& Astronomy, University of Nottingham, NG7 2RD, UK}
\date{\today}% It is always \today, today,

\begin{abstract}
We present a target field approach to analytically design magnetic fields using permanent magnets. We assume that their magnetisation is bound to a two-dimensional surface and is composed of a complete basis of surface modes. By posing the Poisson's equation relating the magnetic scalar potential to the magnetisation using Green's functions, we derive simple integrals which determine the magnetic field generated by each mode. This approach is demonstrated by deriving the governing integrals for optimising axial magnetisation on cylindrical and circular-planar surfaces. We approximate the governing integrals numerically and implement them into a regularised least-squares optimisation routine to design permanent magnets that generate uniform axial and transverse target magnetic fields. The resulting uniform axial magnetic field profiles demonstrate more than a tenfold increase in uniformity across equivalent target regions compared to the field generated by an optimally separated axially magnetised pair of rings, as validated using finite element method simulations. We use a simple example to examine how two-dimensional surface magnetisation profiles can be emulated using thin three-dimensional volumes and determine how many discrete intervals are required to accurately approximate a continuously-varying surface pattern. Magnets designed using our approach may enable higher quality bias fields for electric machines, nuclear fusion, fundamental physics, magnetic trapping, and beyond.
\end{abstract}

\maketitle

\section{\label{sec:intro}Introduction}
Permanent magnets are materials containing iron, cobalt, or nickel that exhibit known stable magnetisation, making them suited for generating static (DC) magnetic fields. The best permanent magnet materials, like ferrite ceramics and neodymium alloys, have a high retentivity, coercivity, and Curie temperature. Thus, their strong retained magnetisation is resistant to being altered by external magnetic fields or temperature changes. Permanent magnets are extensively used in industrial applications to generate static magnetic fields in devices such as electric machines~\cite{978-1-4200-6440-7, 6425118, en14020283, 764898, 8785119} and transducers~\cite{5343533A, US10469951B2}. Moreover, they are an essential component of many scientific experiments and apparatus that require strong static magnetic fields, including nuclear fusion stellerators~\cite{Landreman_2021, LU2022100709, Qian_2022, Xu_2021, Lu_2021}, Magnetic Resonance Imaging (MRI) scanners~\cite{4711303, 8302405}, Zeeman slowers~\cite{10.1119/1.4930080, 10.1063/1.3600897}, ion pumps~\cite{ISOARDI2023111792, LEE2018869}, magnetic atom and ion traps~\cite{PRXQuantum.2.030326, EPFurlani_1997}, and electromagnetic isotope separators~\cite{LIU2021165428}. 

As such, the design of target magnetic field profiles using permanent magnets has garnered significant research interest. Roméo and Hoult's seminal paper in Ref.~\citep{Romeo} established methods for designing target magnetic field profiles by examining symmetries in the magnetic scalar potential and replicating them through the arrangement of simple magnetised structures such as loops and arcs. Drawing on their design principles, analytical models have been developed to create magnetised loops~\cite{PENG2004165} and to study the Fourier series expansions of magnetic fields generated by hypothetical magnetic point charges~\cite{10.1109/TMAG.2004.831653}. Further analytical models have been developed to determine the optimal placement of magnetised rods~\cite{4711303, 8302405} to enhance target field profiles for low-power MRI applications. Alternatively, numerical methods have been developed to solve for the magnetic field generated by magnetised structures with more complex geometries, including surface pixels~\cite{Schmied_2010}, surface meshes~\cite{Landreman_2021, Xu_2021, Lu_2021}, and volumes~\cite{6749021}. While these approaches are very flexible, they provide limited insight into the physical behaviour of permanent magnet systems and often demand extensive computations of integrals or at numerous evaluation points to find precise solutions. Consequently, design optimisation often has a limited resolution and is computationally intensive, restricting the potential for design refinement.

In contrast, when designing current-carrying magnetic field coils, rather than permanent magnets, complex surface patterns may be designed straightforwardly and efficiently using a family of approaches known as \emph{target field methods}, conceived initially by Turner in Ref.~\citep{10.1088/0022-3727/19/8/001}. Target field methods use Green's function solutions to express the magnetic field in terms of Fourier space integrals, with the Fourier transform of surface current flow modes included in the integral argument. By selecting suitable surface flows in an orthogonal basis, typically a Fourier series with symmetries that match those in the Green's function, these integrals can be solved analytically or converge efficiently numerically~\cite{10.1088/0022-3727/34/24/305, 10.1088/0022-3727/35/9/303, 10.1088/0022-3727/36/2/302}. Quadratic optimisation methods, commonly least-squares, can then be used to determine the weights of different current flow modes to achieve the desired target field~\cite{693569}. These current flows may oscillate greatly across the surface to control magnetic field variations and thereby enhance magnetic field fidelity over a target region.

In this paper, we extend the target field method approaches used for designing current flows to the analytic design of surface magnetisation. This enables us to optimise complicated target profiles using complex magnetisation arrangements efficiently and straightforwardly. We begin by employing cylindrical Green’s functions to derive Fourier space integrals that relate axial magnetisation on cylindrical or circular-planar surfaces to the magnetic field via the magnetic scalar potential. We then introduce Fourier series representations of axial magnetisation on these surfaces and input these series representations into the Fourier space integrals. We numerically solve these integrals to design surface magnetisation on cylindrical and circular bi-planar surfaces through least-squares optimisation, regularised by the curvature across the surface of the magnet. These surface magnetisation patterns therefore generate uniform axial and transverse magnetic fields but without excessive spatial oscillations across their surfaces which would increase manufacturing complexity. The magnetic field profiles created by these surface patterns are validated by comparison to Finite Element Method (FEM) simulations, which show strong agreement with theoretical predictions. We also propose approaches to replicate the two-dimensional (2D) surface magnetisation profiles with patterns that can be produced using sets of thin three-dimensional (3D) magnetised regions with incrementally increasing surface magnetisation, such as stacks of equally thin magnetised metal plates. We use a simple example to assess how the thickness of a 3D volume and the number of discrete magnetised regions influences the quality of the generated magnetic field.

\section{\label{sec:model}Mathematical Model}
\subsection{Green's Function Formulation}
Let us consider a region of space containing no electrical currents and only permanent magnets with a fixed magnetisation, $\mathbf{M}$, such that the magnetic flux density, outside the magnets, in free space, is $\mathbf{B}=\mu_0\left(\mathbf{H}+\mathbf{M}\right)$, where $\mathbf{H}$ is the magnetic field strength and $\mu_0$ is the magnetic permeability of free space. In this scenario, the magnetic flux density in free space is simply the gradient of the magnetic scalar potential,
\begin{align}\label{eq.bsh}
    \mathbf{B}\left(\mathbf{r}\right)&=-\mu_0\nabla_{\mathbf{r}}\Psi\left(\mathbf{r}\right),
\end{align}
where the magnetic scalar potential is evaluated at a point $\mathbf{r}=x~\mathbf{\hat{x}}+y~\mathbf{\hat{y}}+z~\mathbf{\hat{z}}$ and $\nabla_{\mathbf{r}}={\partial}/{{\partial}x}~\mathbf{\hat{x}} + {\partial}/{{\partial}y}~\mathbf{\hat{y}} +{\partial}/{{\partial}z}~\mathbf{\hat{z}}$.

Applying Amp\`ere's Law, $\nabla\times\mathbf{H}=0$, and Gauss' Law of magnetism, $\nabla\boldsymbol{\cdot}\mathbf{H} = -\nabla\boldsymbol{\cdot}\mathbf{M}$, then inputting the magnetic scalar potential, \eqref{eq.bsh}, we can generate the following Poisson equation for the magnetic scalar potential in free space,
\begin{equation}\label{eq.Poissiondiff}
    \nabla_{\mathbf{r}}^2\Psi\left(\mathbf{r}\right) = \nabla_{\mathbf{r}}\boldsymbol{\cdot}\mathbf{M}\left(\mathbf{r}\right).
\end{equation}
Defining a Green's function as $\nabla_{\mathbf{r}}^2G(\mathbf{r},\mathbf{r'})=\delta\left(\mathbf{r}-\mathbf{r}'\right)$, where $\delta\left(x\right)$ is the Dirac delta function, we can integrate equation~\eqref{eq.Poissiondiff} to find~\cite{JacksonEMscalar},
\begin{equation}\label{eq.magscalar}
    \Psi\left(\mathbf{r}\right) = -\frac{1}{4\pi} \int \mathrm{d}\mathbf{r'} \ \nabla_{\mathbf{r}} \boldsymbol{\cdot} \left(\frac{\mathbf{M}\left(\mathbf{r'}\right)}  {\left|\mathbf{r}-\mathbf{r}'\right|}\right),
\end{equation}
where the Green's function is now represented as $G(\mathbf{r},\mathbf{r'})={1}/{\left|\mathbf{r}-\mathbf{r}'\right|}$ and $\mathbf{r'}=x'\mathbf{\hat{x}}+y'\mathbf{\hat{y}}+z'\mathbf{\hat{z}}$ is a point within the magnetised region.

Let us now apply the following vector identity,
\begin{multline}\label{eq.vecident_Max}
    \nabla_{\mathbf{r}} \boldsymbol{\cdot} \left( \frac{\mathbf{M}\left(\mathbf{r}'\right)}{\left|\mathbf{r}-\mathbf{r}'\right|}\right) = \frac{1}{\left|\mathbf{r}-\mathbf{r}'\right|} \nabla_{\mathbf{r}} {\boldsymbol{\cdot}} \mathbf{M}\left(\mathbf{r}'\right) \ + \\ \mathbf{M}\left(\mathbf{r}'\right) \nabla_{\mathbf{r}}{\boldsymbol{\cdot}}\left(\frac{1}{\left|\mathbf{r}-\mathbf{r}'\right|}\right),
\end{multline}
and note that the divergence of the magnetisation must be zero outside of the magnetised region, $\nabla_{\mathbf{r}} {\boldsymbol{\cdot}} \mathbf{M}\left(\mathbf{r}'\right)=0$. Substituting equation~\eqref{eq.vecident_Max} into equation~\eqref{eq.magscalar}, we generate an integral representation of the magnetic scalar potential,
\begin{equation}\label{eq.magscalar2}
    \Psi\left(\mathbf{r}\right) = -\frac{1}{4\pi} \int \mathrm{d}\mathbf{r'} \ \mathbf{M}\left(\mathbf{r}'\right) \nabla_{\mathbf{r}}{\boldsymbol{\cdot}} \left(\frac{1}{\left|\mathbf{r}-\mathbf{r}'\right|}\right).
\end{equation}
By imposing that the magnetisation is directed only along the axial direction, $\mathbf{M}=M_z\mathbf{\hat{z}}$, this simplifies further to
\begin{equation}\label{eq.magscalar3}
    \Psi\left(\mathbf{r}\right) = -\frac{1}{4\pi} \int \mathrm{d}\mathbf{r'} \ M_z\left(\mathbf{r}'\right) \frac{\partial}{{\partial}z}\left(\frac{1}{\left|\mathbf{r}-\mathbf{r}'\right|}\right).
\end{equation}

Below, we match the spatial symmetries of the surface upon which we wish to design magnetisation to an appropriate Green's function, and thereby simplify the integral formulation such that the magnetic scalar potential can be determined simply. Then, by expressing the axial magnetisation in terms of mutually orthogonal modes on a surface, the magnetic field can be similarly expressed using corresponding modes that are mutually orthogonal across all of space.

\subsection{Cylindrical Surface}

\begin{figure}[h!]
\centering
\input{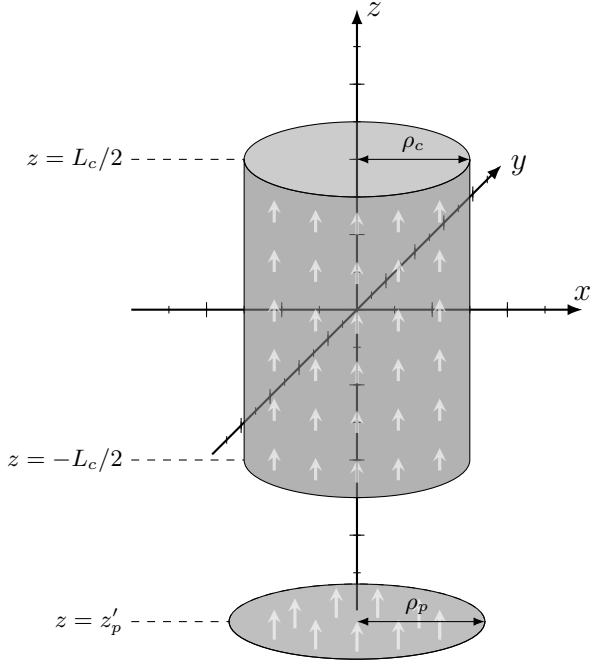}
\caption{Schematic diagram of magnetisation surfaces: a cylinder of radius $\rho_c$ and length $L_c$ and a circular plane of radius $\rho_p$ which lies at axial position $z_p$. White arrows denote the case where the surfaces are axially magnetised along the $+\mathbf{\hat{z}}$ direction.}
\label{fig.surfaces}
\end{figure}
Let us examine axial magnetisation bound to an infinitesimally thin cylinder of radius $\rho_c$ and length $L_c$, centered about the origin, as presented in the upper section of Fig.~\ref{fig.surfaces}. We can represent the axial magnetisation in cylindrical coordinates as
\begin{equation}\label{eq.magsurfacecylinder}
M_z=\delta\left(\rho'-\rho_c\right)\sigma_{zc}\left(\phi',z'\right),
\end{equation}
where $\sigma_{zc}\left(\phi',z'\right)$ is the axial surface magnetisation, i.e. the magnetic dipole moment density per unit area on the cylindrical surface.

Now, we shall apply a Green's function with simple cylindrical symmetries~\cite{10.1007/978-1-4614-9506-2}
\begin{multline}\label{eq.greeNyl}
    \frac{1}{|\textbf{r}-\textbf{r}'|} =\frac{1}{\pi}\sum_{m=-\infty}^{\infty}e^{im(\phi-\phi')} \ \times \\ \int_{-\infty}^{\infty}\mathrm{d}k \ e^{ik\left(z-z'\right)} I_m\left(|k|\rho_<\right)K_m\left(|k|\rho_>\right),
\end{multline}
and examine the magnetic field inside the surface magnetisation, i.e. ${\rho_<}=\rho$ and $\rho_>=\rho'$. Substituting equations~\eqref{eq.greeNyl}~and~\eqref{eq.magsurfacecylinder} into equation~\eqref{eq.magscalar2} and then integrating over the radial coordinate, we find
\begin{multline}\label{eq.magscalar_cyl}
    \Psi\left(\rho,\phi,z\right) = -\frac{i\rho_c}{2\pi}\sum_{m=-\infty}^\infty e^{im\phi} \times \\ \int_{\infty}^{\infty} \mathrm{d}k \ k I_m\left(|k|\rho\right)K_m\left(|k|\rho_c\right)e^{ikz} \sigma_{zc}^m(k),
\end{multline}
where the Fourier transform of the axial magnetisation on the cylindrical surface is
\begin{equation}\label{eq.azft}
    \sigma_{zc}^m(k) = \frac{1}{2\pi}\int_{0}^{2\pi}\mathrm{d}\phi' e^{-im\phi'} \int_{-\infty}^{\infty}\mathrm{d}z'\ e^{-ikz'}\sigma_{zc}\left(\phi',z'\right).
\end{equation}
Thus, the magnetic flux density, expressed in cylindrical coordinates, $\mathbf{B}=B_{\rho}~\boldsymbol{\hat{\rho}} + B_{\phi}~\boldsymbol{\hat{\phi}}+ B_{z}~\boldsymbol{\hat{z}}$, may be formulated by calculating the gradient of the scalar potential, using equation~\eqref{eq.bsh}, with vector components
\begin{multline}\label{eq.Brho_cyl}
    B_{\rho}\left(\rho,\phi,z\right) = -\frac{i\mu_0\rho_c}{2\pi}\sum_{m=-\infty}^\infty  e^{im\phi} \ \times \\ \int_{\infty}^{\infty} \mathrm{d}k \ k |k| I'_m\left(|k|\rho\right)K_m\left(|k|\rho_c\right)e^{ikz} \sigma_{zc}^m(k),
\end{multline}
\begin{multline}\label{eq.Bphi_cyl}
    B_{\phi}\left(\rho,\phi,z\right) = \frac{\mu_0\rho_c}{2\pi\rho}\sum_{m=-\infty}^\infty  me^{im\phi} \ \times \\ \int_{\infty}^{\infty} \mathrm{d}k \ k I_m\left(|k|\rho\right)K_m\left(|k|\rho_c\right)e^{ikz} \sigma_{zc}^m(k),
\end{multline}
\begin{multline}\label{eq.Bz_cyl}
    B_{z}\left(\rho,\phi,z\right) = \frac{\mu_0\rho_c}{2\pi}\sum_{m=-\infty}^\infty e^{im\phi} \ \times \\ \int_{\infty}^{\infty} \mathrm{d}k \ k^2 I_m\left(|k|\rho\right)K_m\left(|k|\rho_c\right)e^{ikz} \sigma_{zc}^m(k).
\end{multline}
The magnetic flux density can then be evaluated by calculating the Fourier transform of the axial surface magnetisation and approximating the Fourier space integrals numerically.

Now, let us return to our magnetised cylinder of radius $\rho_c$ and length $L_c$ and impose a complete basis of orthogonal modes bound to its surface. This is best expressed using a Fourier series,
\begin{multline}\label{eq.Mcyl}
    \sigma_{zc}(\phi',z') = \mathcal{T}\left(z',-L_c/2,L_c/2\right) \sum_{n=1}^N \ \Bigg[ W_{n0} \ + \\ \sum_{m=1}^M\ \left(W_{nm}\cos(m\phi')+Q_{nm}\sin(m\phi')\right) \Bigg]  \ \times \\ \sin \left(\frac{n\pi\left(z'-L_c/2\right)}{L_c}\right),
\end{multline}
with the top-hat operator, $\mathcal{T}\left(x,y,z\right)$, acting to bound the axial magnetisation to the cylindrical surface,
\begin{equation}\label{eq.heaviside_bound}
    \mathcal{T}\left(x,y,z\right) = \mathcal{H}\left(x-y\right)-\mathcal{H}\left(x-z\right),
\end{equation}
where $\mathcal{H}\left(x\right)$ is the Heaviside step function. The symbols $\left(W_{n0},~W_{nm},~Q_{nm}\right)$ denote the weightings of the Fourier modes, hereby referred to as \emph{Fourier coefficients}. This basis is encoded such that the axial surface magnetisation smoothly tends to zero at the axial edge of the cylinder to make the magnet easier to construct.

Substituting equation~\eqref{eq.Mcyl} into equations~\eqref{eq.Brho_cyl}--\eqref{eq.Bz_cyl}, we generate the governing equations
\begin{widetext}
\begin{align}
B_{\rho}\left(\rho,\phi,z\right) &= \sum_{n=1}^{N} \left[ W_{n0}B_{{\rho},c/p}^{n0}\left(\rho,z\right)+\sum_{m=1}^M\left(W_{nm}\cos\left(m\phi\right) + Q_{nm}\sin\left(m\phi\right)\right)B_{{\rho},c/p}^{nm}\left(\rho,z\right) \right], \\ B_{\phi}\left(\rho,\phi,z\right) &= \sum_{n=1}^{N} \sum_{m=1}^M\left(-W_{nm}\sin\left(m\phi\right) + Q_{nm}\cos\left(m\phi\right)\right)B_{{\phi},c/p}^{nm}\left(\rho,z\right), \\
B_{z}\left(\rho,\phi,z\right) &= \sum_{n=1}^{N} \left[ W_{n0}B_{z,c/p}^{n0}\left(\rho,z\right)+\sum_{m=1}^M\left(W_{nm}\cos\left(m\phi\right) + Q_{nm}\sin\left(m\phi\right)\right)B_{z,c/p}^{nm}\left(\rho,z\right) \right],
\end{align}
\end{widetext}
where $\left(B_{{\rho},c/p}^{nm},~B_{{\phi},c/p}^{nm},~B_{{z},c/p}^{nm}\right)$, represent the governing integrals that weight the Fourier coefficients with the subscripts $c$ and $p$ denoting the cylindrical and circular-planar cases, respectively [see subsection~\ref{ssec:circular}]. In the cylindrical case, these governing integrals are
\begin{align}\label{eq.cyl-start}
B_{{\rho},c}^{nm}\left(\rho,z\right)&=2n\mu_0\rho_c L_c \int_{0}^{\infty}\mathrm{d}k\ \times \nonumber \\ &\hspace{-30pt}\frac{k^2I'_m\left(k\rho\right)K_m\left(k\rho_c\right)}{n^2\pi^2-L_c^2k^2}
     \begin{Bmatrix}
     \sin\left(kz\right)\cos\left(kL_c/2\right) \\ 
     -\cos\left(kz\right)\sin\left(kL_c/2\right)
     \end{Bmatrix},
\end{align}
\begin{align}
B_{{\phi},c}^{nm}\left(\rho,z\right)&=\frac{2nm\mu_0\rho_cL_c}{\rho} \int_{0}^{\infty}\mathrm{d}k\ \times \nonumber \\ &\hspace{-30pt}\frac{kI_m\left(k\rho\right)K_m\left(k\rho_c\right)}{n^2\pi^2-L_c^2k^2}
     \begin{Bmatrix}
     \sin\left(kz\right)\cos\left(kL_c/2\right) \\ 
     -\cos\left(kz\right)\sin\left(kL_c/2\right)
     \end{Bmatrix},
\end{align}
\begin{align}\label{eq.cyl-end}
    B_{z,c}^{nm}\left(\rho,z\right)&= 2n\mu_0\rho_cL_c \int_{0}^{\infty}\mathrm{d}k\ \times \nonumber \\ &\hspace{-30pt}\frac{k^2I_m\left(k\rho\right)K_m\left(k\rho_c\right)}{n^2\pi^2-L_c^2k^2}
     \begin{Bmatrix}
     \cos\left(kz\right)\cos\left(kL_c/2\right) \\ 
     \sin\left(kz\right)\sin\left(kL_c/2\right)
     \end{Bmatrix},
\end{align}
where the vector $\{ 2n-1, \ 2n \}$ denotes the magnetic flux density for either odd, $2n-1$, or even, $2n$, orders for $n\in\mathbb{Z}^+$.

These integrals can be approximated numerically to determine the magnitude of the field generated by each axial surface magnetisation mode at each point in space. During numerical approximation, the high spatial frequency limit, $k\to\infty$, in the integrands may be approximated as a large number, as determined by the spatial frequency of the oscillations (see example in section~\ref{sec:results}). Designs on a larger geometry oscillate more at a given spatial frequency and so the integrals converge faster.

\subsection{Circular-Planar Surface}\label{ssec:circular}
Now, let us examine the magnetic field above or below a source of axial magnetisation bound to a circular ${\rho}{\phi}$-plane of radius $\rho_p$ at axial position $z'=z_p$, as presented in the lower section of Figure~\ref{fig.surfaces}. We shall express this surface magnetisation as
\begin{equation}\label{eq.magsurfaceplane}
M_z=\delta\left(z'-z_p\right)\sigma_{zp}\left(\rho',\phi'\right),
\end{equation}
where $\sigma_{zp}\left(\rho',\phi'\right)$ is the axial surface magnetisation bound to the circular plane. 

To exploit the radial symmetries in this system, we shall apply a Green's function in cylindrical coordinates defined in terms of Bessel functions of the first kind~\cite{10.1007/978-1-4614-9506-2},
\begin{multline}\label{eq.green1}
    G(\mathbf{r,r'}) =\sum_{m=-\infty}^{\infty}e^{im(\phi-\phi')} \times \\ \int_{0}^{\infty}\mathrm{d}k \ e^{-k\left|z-z'\right|}
     J_m(k\rho)J_m(k\rho').
\end{multline}
Substituting equation~\eqref{eq.green1} into equation~\eqref{eq.magscalar2}, and then integrating over the axial coordinate, we find
\begin{multline}\label{eq.magscalar_planar}
    \Psi\left(\rho,\phi,z\right) = \frac{1}{2}\frac{\left(z-z_p\right)}{\left|z-z_p\right|}\sum_{m=-\infty}^\infty e^{im\phi} \times \\ \int_{0}^{\infty} \mathrm{d}k \ ke^{-k\left|z-z_p\right|} J_m\left(k\rho\right) \sigma_{zp}^m(k).
\end{multline}
Here, we define $\sigma_{zp}^m(k)$ as the cylindrical Fourier transform of the axial surface magnetisation,
\begin{equation}\label{eq.ht}
    \sigma_{zp}^m(k)=\frac{1}{2\pi}\int_0^\infty{d}\rho'\rho'\ \int_0^{2\pi}\mathrm \mathrm{d}\phi' \ e^{-im\phi'}J_m(k\rho')\sigma_{zp}(\rho',\phi').
\end{equation}
Equation~\eqref{eq.ht} is sometimes referred to as the Hankel transform of the $m^{\textnormal{th}}$ order. Substituting equation~\eqref{eq.magscalar_planar} into equation~\eqref{eq.bsh}, the magnetic flux density vector components are
\begin{multline}\label{eq.Brho_plan}
    B_{\rho}\left(\rho,\phi,z\right) = \frac{\mu_0}{2}\frac{\left(z-z_p\right)}{\left|z-z_p\right|}\sum_{m=-\infty}^\infty  e^{im\phi} \ \times \\ \int_{0}^{\infty} \mathrm{d}k \ k^2 e^{-k\left|z-z_p\right|} J'_m\left(k\rho\right) \sigma_{zp}^m(k),
\end{multline}
\begin{multline}\label{eq.Bphi_plan}
    B_{\phi}\left(\rho,\phi,z\right) = \frac{i\mu_0}{2\rho}\frac{\left(z-z_p\right)}{\left|z-z_p\right|}\sum_{m=-\infty}^\infty  me^{im\phi} \ \times \\ \int_{0}^{\infty} \mathrm{d}k \ k e^{-k\left|z-z_p\right|} J_m\left(k\rho\right) \sigma_{zp}^m(k),
\end{multline}
\begin{multline}\label{eq.Bz_plan}
    B_{z}\left(\rho,\phi,z\right) = -\frac{\mu_0}{2}\sum_{m=-\infty}^\infty  e^{im\phi} \ \times \\ \int_{0}^{\infty} \mathrm{d}k \ k^2 e^{-k\left|z-z_p\right|} J_m\left(k\rho\right) \sigma_{zp}^m(k).
\end{multline}

A complete basis of orthogonal modes on a circular surface can be expressed in terms of a Fourier--Bessel series, which is the discrete counterpart of the Hankel transform. Thus, we represent the axial surface magnetisation bound to the circular plane as
\begin{multline}\label{eq.mag_circ}
    \sigma_{zp}(\rho',\phi')=\mathcal{T}\left(\rho',0,\rho_p\right) \sum_{n=1}^N \Bigg[ W_{n0} J_{0}\left(\frac{\varrho_{n0}\rho'}{\rho_p}\right) \ + \\ \sum_{m=1}^M\ \left(W_{nm}\cos(m\phi')+Q_{nm}\sin(m\phi')\right) J_{m}\left(\frac{\varrho_{nm}\rho'}{\rho_p}\right) \Bigg],
\end{multline}
where $\varrho_{nm}$ is the n\textsuperscript{th} zero of the Bessel function of the first kind of order $m$ such that $J_m\left(\varrho_{nm}\right)=0$.

Substituting equation~\eqref{eq.mag_circ} into equations~\eqref{eq.Brho_plan}--\eqref{eq.Bz_plan} and integrating over the surface of the magnet, we find that the governing equations are of the same form as equations~\eqref{eq.Brho_cyl}--\eqref{eq.Bz_cyl} with new integral weightings,
\begin{align}\label{eq.planar-start}            
    B_{{\rho},p}^{nm}\left(\rho,z\right)&=\frac{\mu_0\varrho_{nm}\rho_p^2J'_m\left(\varrho_{nm}\right)}{2} \frac{\left(z-z_p\right)}{\left|z-z_p\right|} \ \times \nonumber \\ &\hspace{0pt} \int_{0}^{\infty}\mathrm{d}k\ \frac{k^2J'_m\left(k\rho\right)J_m\left(k\rho_p\right)e^{-k\left|z-z_p\right|}}{\varrho_{nm}^2-k^2\rho_p^2},
\end{align}
\begin{align}
    B_{{\phi},p}^{nm}\left(\rho,z\right)&=\frac{\mu_0 m \varrho_{nm}\rho_p^2J'_m\left(\varrho_{nm}\right)}{2\rho}\frac{\left(z-z_p\right)}{\left|z-z_p\right|} \ \times \nonumber \\ &\hspace{0pt} \int_{0}^{\infty}\mathrm{d}k\ \frac{kJ_m\left(k\rho\right)J_m\left(k\rho_p\right)e^{-k\left|z-z_p\right|}}{\varrho_{nm}^2-k^2\rho_p^2},
\end{align}
\begin{align}\label{eq.planar-end}
    B_{z,p}^{nm}\left(\rho,z\right)&=-\frac{\mu_0\varrho_{nm}\rho_p^2J'_m\left(\varrho_{nm}\right)}{2} \ \times \\ &\hspace{0pt} \int_{0}^{\infty}\mathrm{d}k\ \frac{k^2J_m\left(k\rho\right)J_m\left(k\rho_p\right) e^{-k\left|z-z_p\right|}} {\varrho_{nm}^2-k^2\rho_p^2}.
\end{align}

When designing surface magnetisation on multiple co-axial circular planes, we assume that their Fourier modes are spatially orthogonal and adjust the radius and axial position in equations~\eqref{eq.planar-start}--\eqref{eq.planar-end} based on the geometry of each plane.

\subsection{Optimisation}
Now, we shall use the target integrals, equations~\eqref{eq.cyl-start}--\eqref{eq.cyl-end} and \eqref{eq.planar-start}--\eqref{eq.planar-end} in the cylindrical and circular-planar cases, respectively, to determine the optimal Fourier coefficients, $\left(W_{n0},~W_{nm},~Q_{nm}\right)$, to generate a target magnetic field profile. Here, for simplicity and computational speed, we implement this using least-squares optimisation of a quadratic cost function, however, other more sophisticated techniques may be applied to the same objective integrals, including linear optimisation with inequality constraints~\cite{mkinen2020magneticfield,zetter2020magneticfield}. 

Let us pose a quadratic cost function,
\begin{equation}\label{eq.functional}
f\left(W_n,~W_{nm},~Q_{nm}\right)=\sum_k^{N^\mathrm{target}}\ {\Delta}\mathbf{B}\left(\mathbf{r}_k\right)^2 + \beta C,
\end{equation}
which is determined by the square of the deviation,
\begin{equation}
    {\Delta}\mathbf{B}\left(\mathbf{r}\right) = \mathbf{B}^{\mathrm{target}}\left(\mathbf{r}\right)-\mathbf{B}\left(W_n,~W_{nm},~Q_{nm};\mathbf{r}\right),
\end{equation}
of the modal magnetic flux density from a target, $\mathbf{B}^{\mathrm{target}}\left(\mathbf{r}\right)$,
evaluated at $k\in[1,N^\mathrm{target}]$ points within a target field region.

The cost function in equation~\eqref{eq.functional} includes a regularisation parameter, $C$, weighted by a parameter $\beta$, that is quadratic with respect to the Fourier coefficients. Here, we set the regularisation to be the curvature of the axial surface magnetisation,
\begin{equation} \label{eq.crvnew}
    C=\int_{r'}\mathrm{d}^2\mathbf{r}'\ |\mathbf{\nabla}^2M_z(\mathbf{r}')|^2.
\end{equation}
We choose to regularise using curvature because it relates closely to how easy it is to manufacture the designs. This means we can control design manufacturability with respect to magnetic field quality by changing the regularisation parameter and recalculating the Fourier coefficients. Substituting the cylindrical axial surface magnetisation, \eqref{eq.Mcyl}, into equation~\eqref{eq.crvnew}, calculating its Laplacian, and then integrating over the surface of the cylinder, we find that the curvature of the cylindrical surface magnetisation is
\begin{multline}
       C=\pi{\rho_c}{L_c} \sum_{n=1}^N\ \Bigg[ \frac{n^4\pi^4}{L_c^4} W_{n0}^2 \ + \\ \frac{1}{2}\sum_{m=1}^M\ \left(\frac{{m^2}}{\rho_c^2}+\frac{n^2\pi^2}{L_c^2}\right)^2 \left(W_{nm}^2+Q_{nm}^2\right)\Bigg].
\end{multline}
Following the same approach, we find that the curvature of the circular-planar surface magnetisation, \eqref{eq.mag_circ}, is
\begin{multline}
         C= \frac{1}{2\rho_p^2}\sum_{n=1}^N \Bigg[2\varrho_{n0}^4 J'_0\left(\varrho_{n0}\right)^2 W_{n0}^2 \ + \\ \sum_{m=1}^M \varrho_{nm}^4 J'_m\left(\varrho_{nm}\right)^2 \left(W_{nm}^2 + Q_{nm}^2\right)\Bigg].
\end{multline}

The optimisation proceeds by finding values of Fourier coefficients which make the derivative of the cost function, \eqref{eq.functional}, with respect to the Fourier coefficients, equal to zero~\cite{10.1007/3-540-31720-1},
\begin{equation}\label{eq.mindif}
    \frac{\partial f}{\partial W_{i0}}=0, \qquad \frac{\partial f}{\partial W_{ij}}=0, \qquad \frac{\partial f}{\partial Q_{ij}}=0.
\end{equation}
These Fourier coefficients minimise deviations from the target field across the target points as well as the curvature of the axial magnetisation. We can then adjust the number and locations of the target points and the weighting of the regularisation to determine the best possible design for a given scenario.

\begin{figure*}[!htb]
\centering
\begin{tabular}{c c c}
\includegraphics[width=0.675\columnwidth]{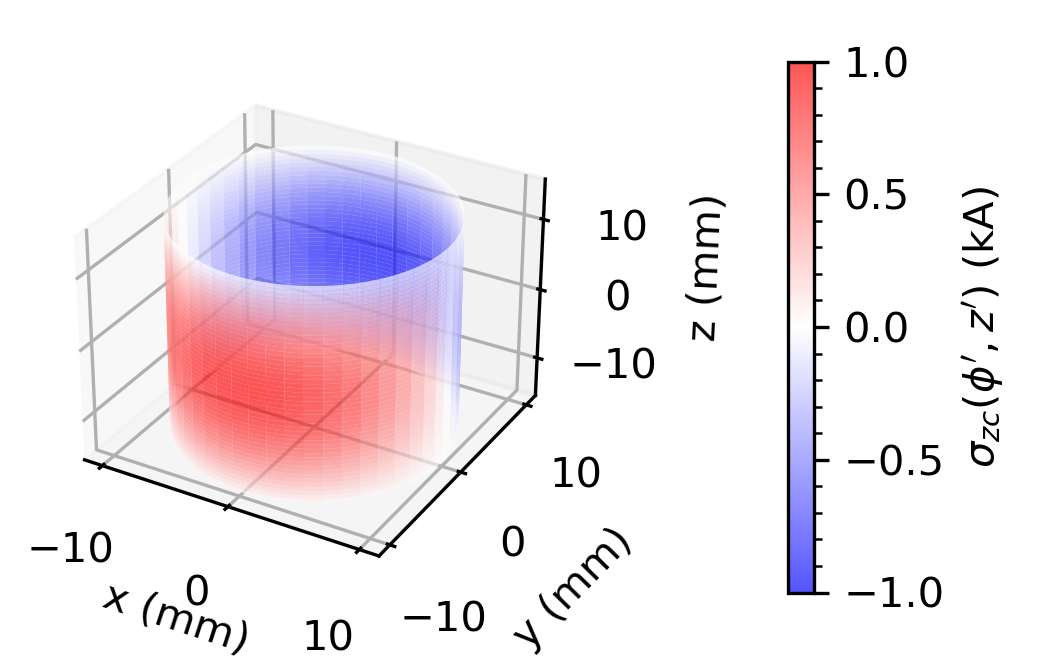} & 
\includegraphics[width=0.675\columnwidth]{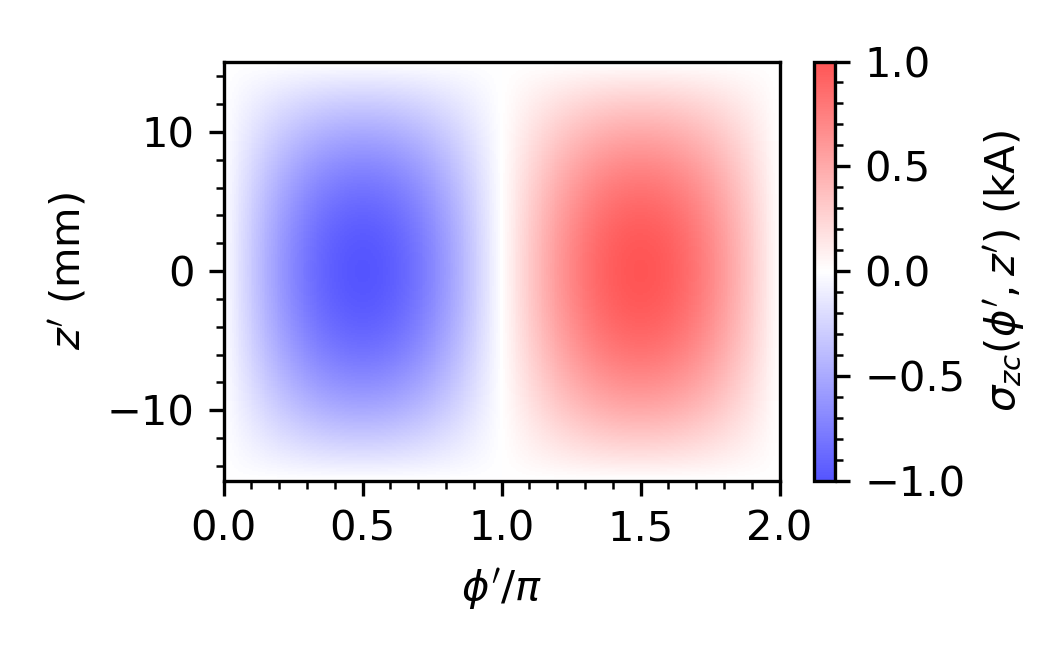} &
\includegraphics[width=0.675\columnwidth]{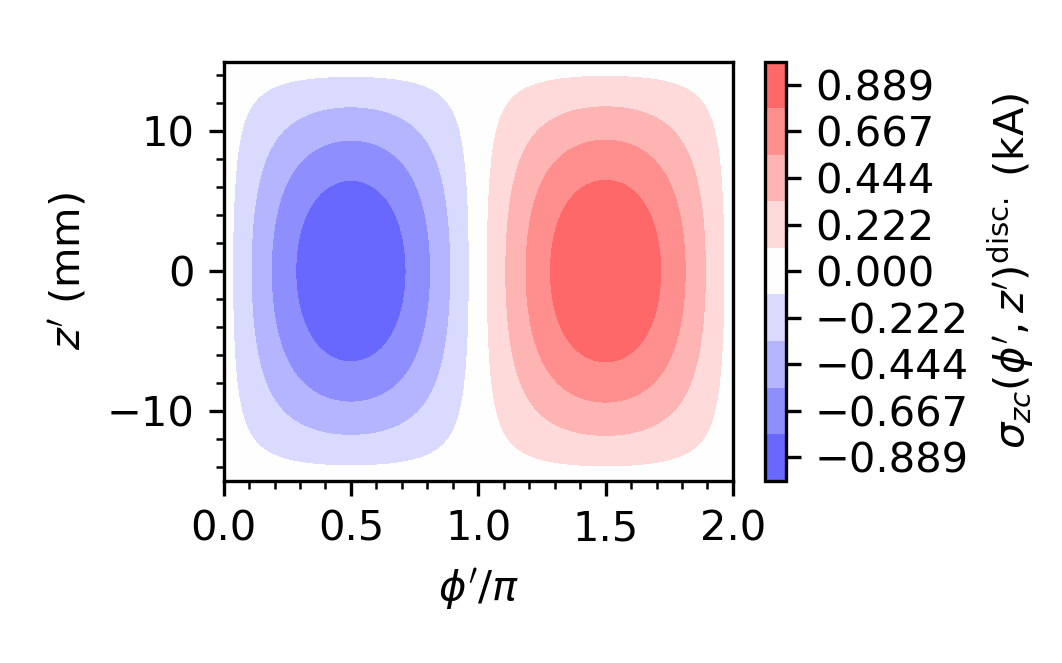} \\
\small{(a)} & \small{(b)} & \small{(c)}
\end{tabular}
\caption{Axial surface magnetisation from equation~\eqref{eq.Mcyl}, for $Q_{1,1}=1\times10^3$~A and all other Fourier coefficients zero, on a cylindrical surface of radius $\rho_c=10$~mm and length $L_c=30$~mm: (a) Presented in 3D and (b) presented in 2D for the surface continuum and in (c) for $N^\mathrm{disc.}=9$ levels. Red/blue colouring indicates positive/negative axial magnetisation with increased intensity representing greater strength.}
\label{fig.contouring}
\end{figure*}
\subsection{Surface Pattern Discretisation}
After determining the optimal Fourier coefficients via least-squares optimisation, we substitute them into equations~\eqref{eq.Mcyl} [cylinder] or~\eqref{eq.mag_circ} [circular plane] to calculate the optimal surface magnetisation patterns. The resulting optimal patterns feature continuously varying magnetisation across their surfaces, which could be challenging to manufacture. To address this, we shall now demonstrate a process for approximating the surface magnetisation at a set of discrete values, where each continuous value is substituted with its closest discrete counterpart. 

Let us demonstrate this approach for the cylindrical case. We evaluate an odd number of equally spaced domain levels, $N^\mathrm{disc.} = 2n + 1$ for $n \in \mathbb{Z}^{+}$, and assume that all surface magnetisation within a specific bound of each level is equal to the value at its centrepoint. The domain levels are centered at discrete axial surface magnetisation values
\begin{equation}
\sigma_{zc}^\mathrm{disc.}=n^\mathrm{disc.}\Delta^\mathrm{disc.}_{zc},
\end{equation}
where the span of each domain level is
\begin{equation}
\Delta^\mathrm{disc.}_{zc} = \mathrm{max}\left(2\left|\sigma_{zc}\right|\right)/N^\mathrm{disc.}.
\end{equation}
Each discrete level encompasses all the continuous values between $\sigma_{zc} = \sigma_{zc}^\mathrm{disc.} + [-\Delta^\mathrm{disc.}_{zc}/2, \Delta^\mathrm{disc.}_{zc}/2]$. The level indices are evenly distributed half-integers,
\begin{multline}
n^\mathrm{disc.}=[-N^\mathrm{disc.}/2+1/2,-N^\mathrm{disc.}/2+3/2,\ldots, \\ N^\mathrm{disc.}/2-3/2,N^\mathrm{disc.}/2-1/2].
\end{multline}

An example of this process for a single Fourier mode is illustrated in Figure~\ref{fig.contouring}. Unlike contouring approaches typically used in magnetic field coil design~\cite{10.1016/j.jmr.2010.09.002}, the domains are evenly distributed around the origin with one level always set at $\sigma_{zc}^\mathrm{disc.} = 0$, where no magnetic material is present (see white zero-level in Fig.~\ref{fig.contouring}c). This ensures that the discrete pattern can be represented by expanding the 2D surface into a thin 3D volume with incrementally varying thickness.

In the varying thickness scenario, instead of an infinitely thin cylinder, the magnetisation can be approximated as being distributed on a thin cylindrical tube centered at radial distance $\rho_c$ with a fixed axial magnetisation $M_{0z}$ and a small radial thickness $\rho_{c0} << \rho_c$. The radial wall thickness of each domain is now
\begin{equation}
\rho^\mathrm{disc.}_{c0}=\left|n^\mathrm{disc.}\right|\Delta^\mathrm{disc.}_{zc}/M_{0z}.
\end{equation}
The polarity (positive or negative) of the axial magnetisation is determined by $\mathrm{sgn}(n^\mathrm{disc.})$, where $\mathrm{sgn}$ is $+1$ if the argument is positive and $-1$ if the argument is negative. 

In principle, these patterns could be constructed using cylindrical permanently axially magnetised plates of equal thicknesses overlaid on top of one another. The number of plates at each point on the surface would be equal to $\left|n^\mathrm{disc.}\right|$. If the number of domain levels $N^\mathrm{disc.}$ is insufficient or the maximum tube thickness $\max(\rho^\mathrm{disc.})$ increases compared to $\rho_c$, errors may be introduced into the model. This can be analysed \emph{a posteriori}, as demonstrated in section~\ref{ssec:analysis} for a simple case study.

This approach can also be applied to the circular-planar example. Instead of extruding the cylinder along the radial direction, we extrude the plane along the axial direction. The approach remains the same, except that the plates are now placed on a circular plane, centered at axial position $z_p$, with varying axial thickness
\begin{equation}
z^\mathrm{disc.}_{p0}=\left|n^\mathrm{disc.}\right|\Delta^\mathrm{disc.}_{zp}/M_{0z},
\end{equation}
where
\begin{equation}
    \Delta^\mathrm{disc.}_{zp} = \mathrm{max}\left(2\left|\sigma_{zp}\right|\right)/N^\mathrm{disc.}.
\end{equation}

\section{\label{sec:results}Results}
Now, we use the mathematical model to design magnetic field-generating arrangements which generate uniform axial, $B_z$, and transverse $B_y$, magnetic fields. We shall design four of these arrangements on independent axially magnetised cylinders and circular bi-planes.

\begin{figure}[ht]
\centering
\includegraphics[width=0.9\columnwidth]{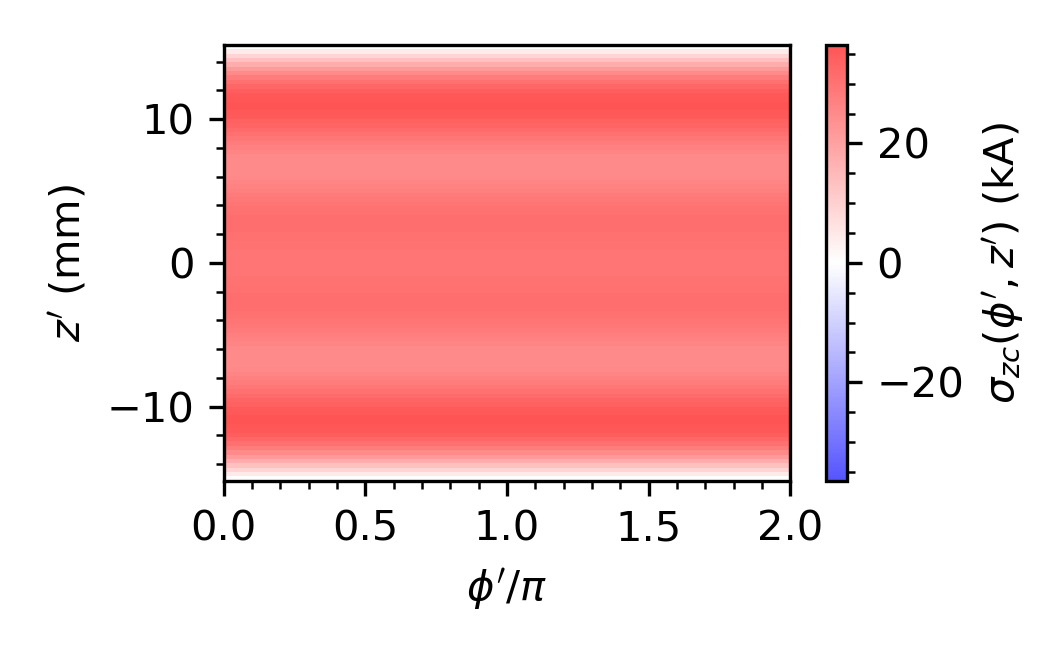} \\ \small{(a)} \\ 
\includegraphics[width=0.9\columnwidth]{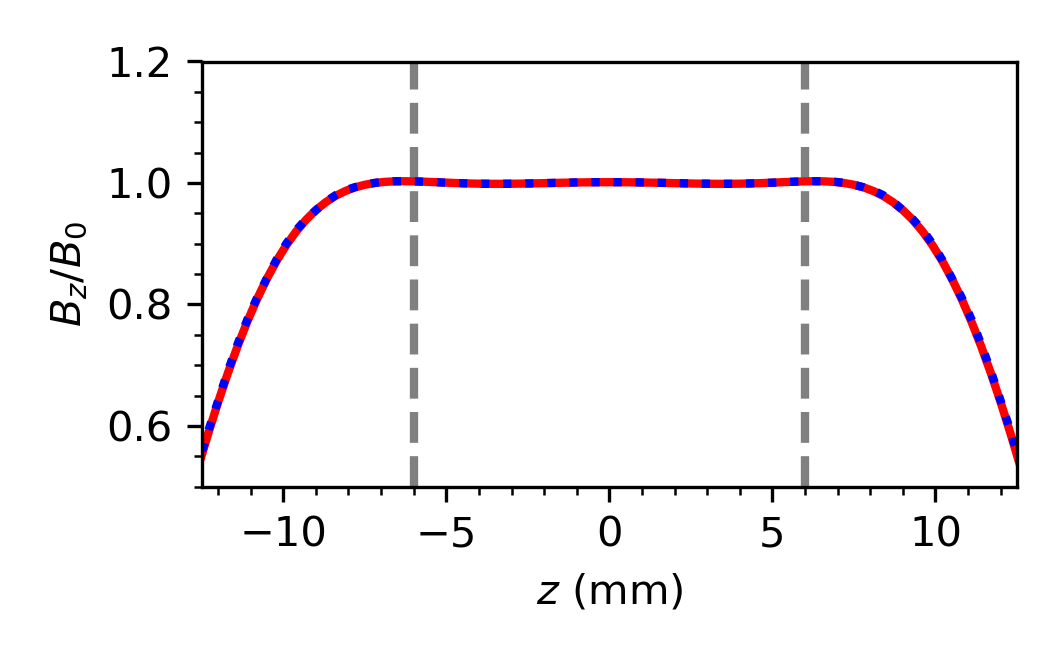} \\
\small{(b)}
\caption{(a) Axial surface magnetisation, $\sigma_{zc}$, on a cylinder of radius $\rho_c=10$~mm and length $L_c=30$~mm. Red/blue colouring indicates positive/negative axial magnetisation with increased intensity representing greater strength. (b) Normalised axial magnetic field, $B_z/B_0$, where $B_0=10$~mT, along the $z$-axis calculated from the analytic model [red solid] and simulated using FEM [blue dotted], where the cylinder is of thickness $\rho_{c0}=1$~mm. The edges of the target region are highlighted [dashed grey lines].}
\label{fig.B10_continuum_cylindrical}
\end{figure}
Let us first consider a cylindrical surface of radius $\rho_c=10$~mm and length of $L_c=30$~mm, which is co-axial and co-centered with the origin. In Fig.~\ref{fig.B10_continuum_cylindrical}a, we present an axial surface magnetisation pattern designed to generate a uniform axial magnetic field, $B_{0}=10$~mT, within a target region extending over $40$\% of the cylinder's length, from $z = [-6,6]$~mm. The pattern is composed of $N = 50$, $M=0$ Fourier modes with $W_{n,0}$ Fourier coefficients. The governing integrals, \eqref{eq.cyl-start}--\eqref{eq.cyl-end}, converge to an accuracy above one part in $10^6$ using a maximum Fourier spatial frequency, $k^\mathrm{max.} = 1 \times 10^4$. The integrals are approximated numerically using the \texttt{scipy.integrate.quad()} function in Python, taking $20.8$ seconds to compute for all $50$ modes at $N^\mathrm{target}=120$ points~\footnote{Calculations are bench-marked using a \textit{MacBookPro18,2} equipped with a \textit{M1~Max} processor with eight 3228~MHz \textit{performance} cores, two 2064~MHz \textit{efficiency} cores, and 32~GB of 512-bit LPDDR5 SDRAM memory.}. The least-squares optimisation, performed using \texttt{numpy.linalg.lstsq()}, takes less than $0.1$ seconds.

In Fig.~\ref{fig.B10_continuum_cylindrical}b, the axial field generated by this pattern along the $z$-axis, calculated using equation~\eqref{eq.Bz_cyl}, is compared to FEM simulations of the axial magnetisation, implemented using COMSOL Multiphysics\textsuperscript{\textregistered} 6.2a. The simulations approximate the axial surface magnetisation as an axial volume magnetisation distributed across a cylinder with a thickness of $\rho_{c0} = 1$ mm. The deviations between the analytical model and FEM simulations are minor, within $<0.05$\% in the target region, validating the design model. These small discrepancies stem from approximating the surface magnetisation as a volume magnetisation and FEM meshing limitations. A large mesh resolution helps to account for greater spatial gradients from higher-order modes in equation~\eqref{eq.Mcyl}. Consequently, the error increases in regions with higher field gradients, such as outside the target region.

\begin{figure}[ht]
\centering
\includegraphics[width=0.9\columnwidth]{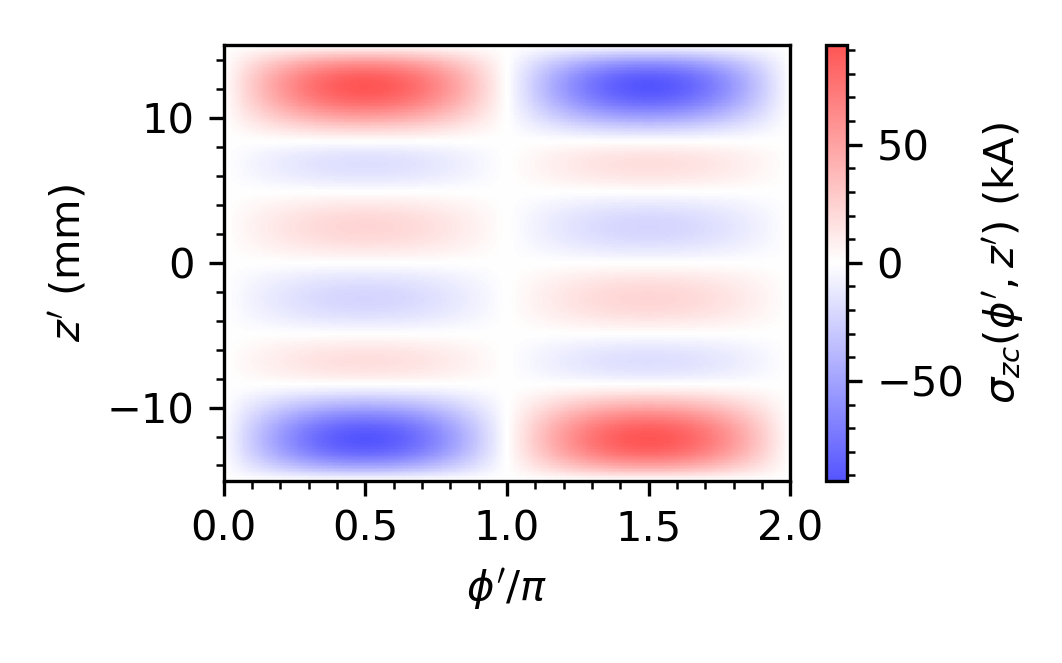} \\ \small{(a)} \\ 
\includegraphics[width=0.9\columnwidth]{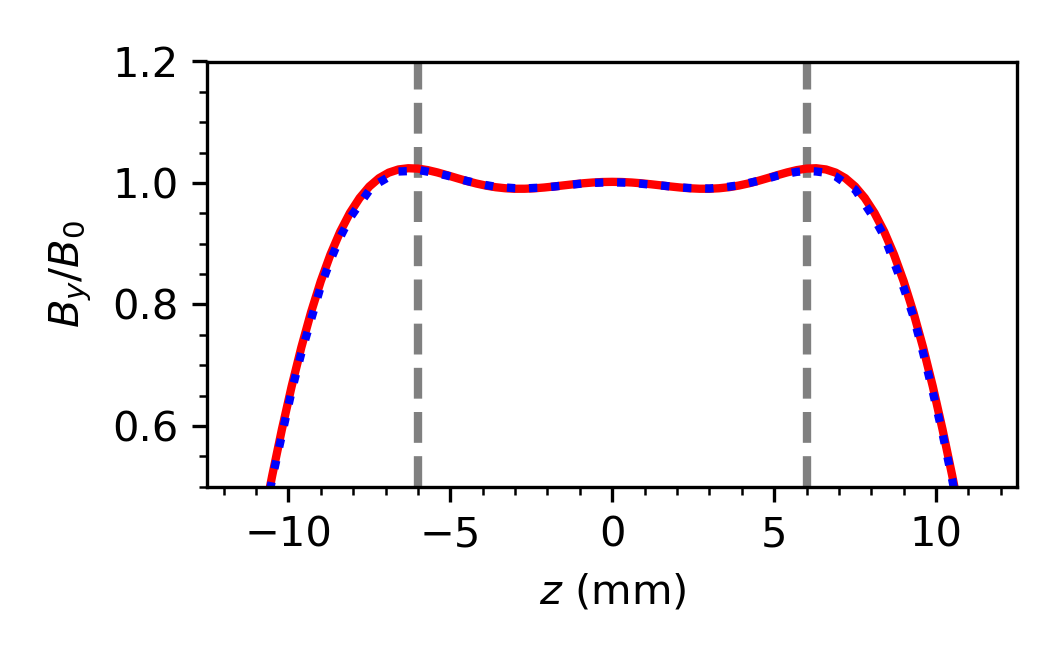} \\
\small{(b)}
\caption{(a) Axial surface magnetisation, $\sigma_{zc}$, on a cylinder of radius $\rho_c=10$~mm and length $L_c=30$~mm. Red/blue colouring indicates positive/negative axial magnetisation with increased intensity representing greater strength. (b) Normalised transverse magnetic field, $B_y/B_0$, where $B_0=10$~mT, along the $z$-axis calculated from the analytic model [red solid] and simulated using FEM [blue dotted], where the cylinder is of thickness $\rho_{c0}=1$~mm. The edges of the target region are highlighted [dashed grey lines].}
\label{fig.B11_continuum_cylindrical}
\end{figure}
In Fig.~\ref{fig.B11_continuum_cylindrical}a, we present an axial surface magnetisation pattern designed to generate a uniform $B_y$ field (transverse to the cylinder's axis) on the same cylindrical surface. The resulting transverse magnetic field along the $z$-axis is calculated and compared with FEM simulation results in Fig.~\ref{fig.B11_continuum_cylindrical}b. The agreement between the analytical model and FEM simulations remains within $<0.05\%$ in the target region. However, the intrinsic uniformity of the transverse field, which exhibits maximum field errors of $<2\%$ in the target region, is diminished compared to the axial field-generating case above. The field is less uniform because the transverse design requires a more oscillatory surface pattern compared to the axial design (seen when comparing Figs.~\ref{fig.B10_continuum_cylindrical}a and \ref{fig.B11_continuum_cylindrical}a qualitatively). Even though the resulting design oscillates more, the curvature regularisation still needs to be weighted more significantly in the least-squares optimisation of the uniform transverse field, which diminishes the magnetic field uniformity.

\begin{figure}[ht]
\centering
\includegraphics[width=0.9\columnwidth]{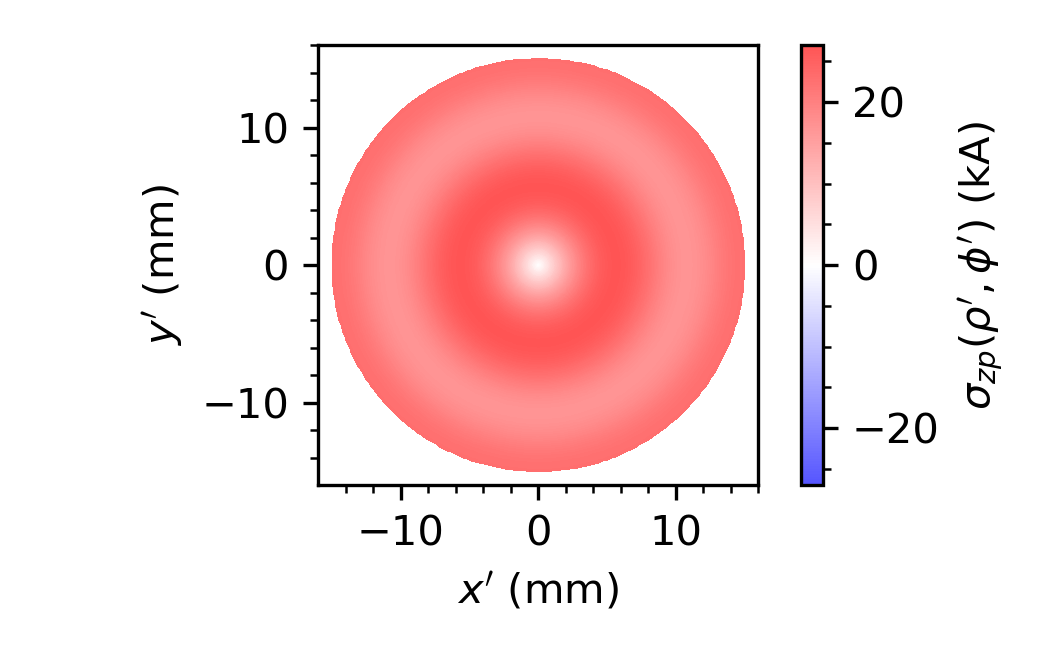} \\ \small{(a)} \\ 
\includegraphics[width=0.9\columnwidth]{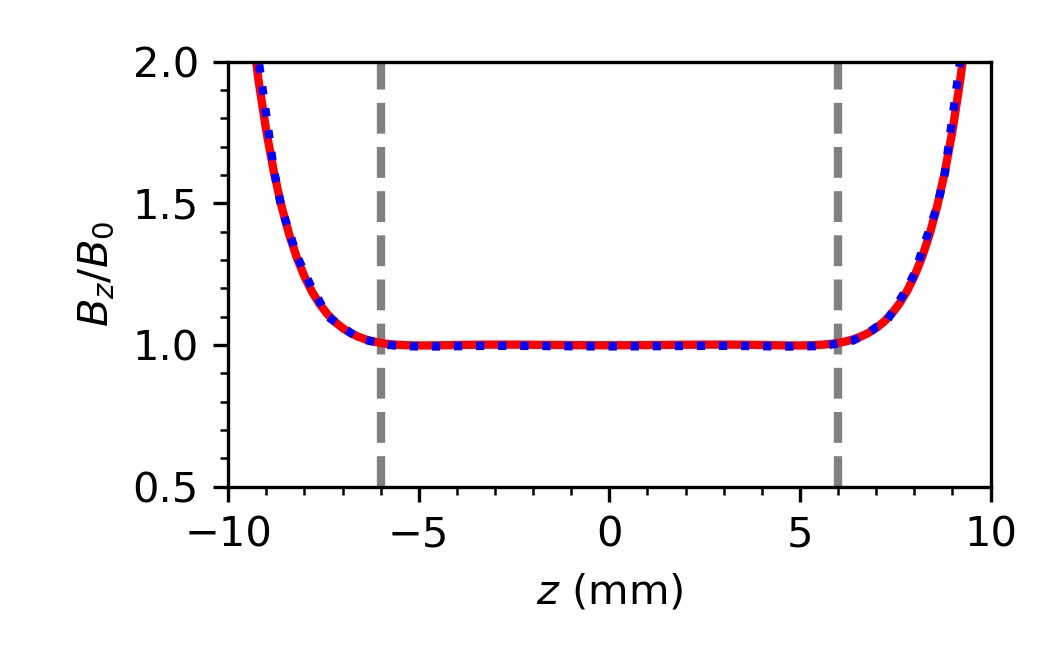} \\
\small{(b)}
\caption{(a) Axial surface magnetisation, $\sigma_{zp}$, on the upper circular plane of radius $\rho_p=15$~mm of an axially magnetised bi-planar pair at axial positions $z={\pm}10$~mm. Red/blue colouring indicates positive/negative axial magnetisation with increased intensity representing greater strength. The magnetisation pattern on the lower plane is identical to that on the upper plane. (b) Normalised axial magnetic field, $B_z/B_0$, where $B_0=10$~mT, along the $z$-axis calculated from the analytic model [red solid] and simulated using FEM [blue dotted], where the planes are of thickness $z_{c0}=1$~mm. The edges of the target region are highlighted [dashed grey lines].}
\label{fig.B10_continuum_planar}
\end{figure}
\begin{figure}[ht]
\centering
\includegraphics[width=0.9\columnwidth]{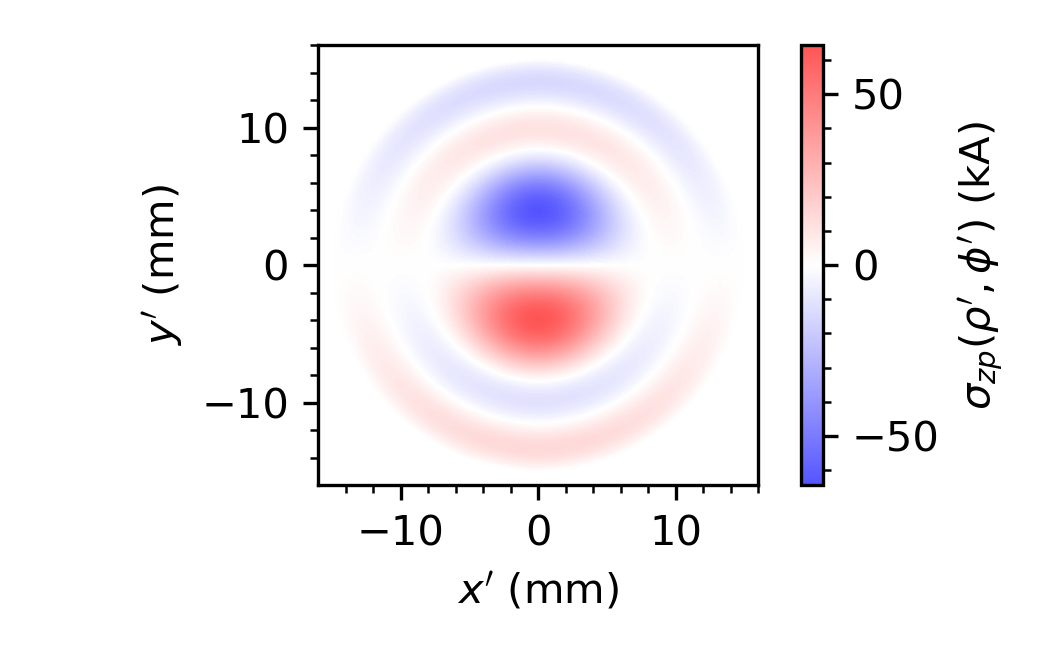} \\ \small{(a)} \\ 
\includegraphics[width=0.9\columnwidth]{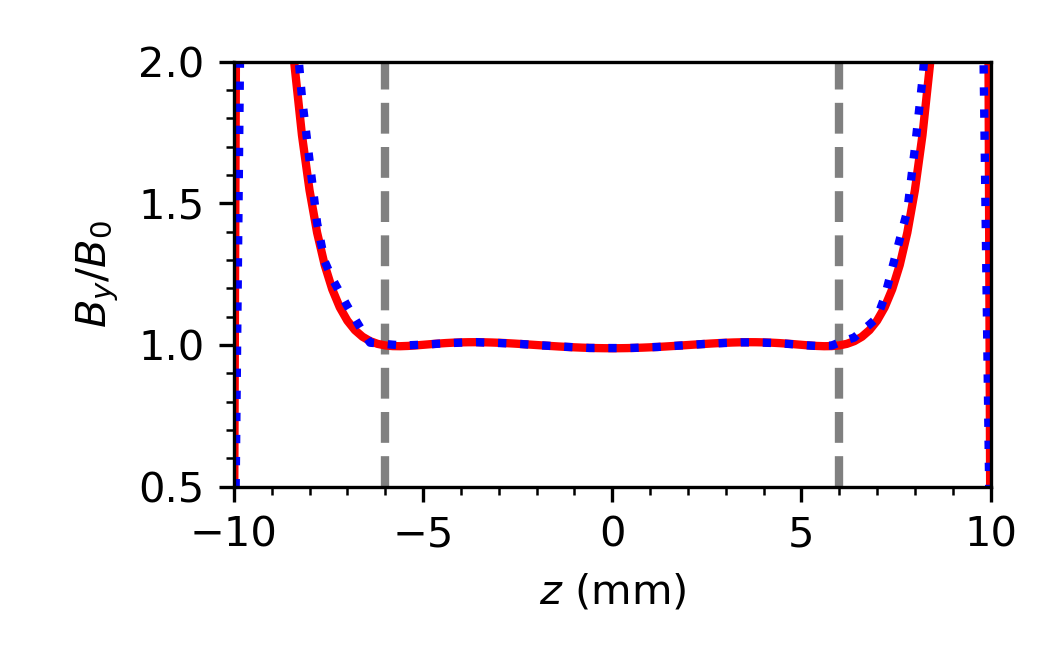} \\
\small{(b)}
\caption{(a) Axial surface magnetisation, $\sigma_{zp}$, on the upper circular plane of radius $\rho_p=15$~mm of an axially magnetised bi-planar pair at axial positions $z={\pm}10$~mm. The magnetisation pattern on the lower plane is equal and opposite to that on the upper plane. (b) Normalised transverse magnetic field, $B_y/B_0$, where $B_0=10$~mT, along the $z$-axis calculated from the analytic model [red solid] and simulated using FEM [blue dotted], where the planes are of thickness $z_{c0}=1$~mm. The edges of the target region are highlighted [dashed grey lines].}
\label{fig.B11_continuum_planar}
\end{figure}
We now design configurations to generate equivalent magnetic fields using a pair of circular bi-planar surfaces instead of cylindrical surfaces. Designing on bi-planes is generally preferred when the system's aspect ratio (length-to-diameter) is less than one~\cite{PhysRevApplied.15.064006}. To compare with the cylindrical case, we therefore design arrangements on a pair of axially magnetised circular bi-planes with a radius of $\rho_p = 15$~mm, positioned at $z = \pm z_p$ where $z_p = 10$~mm. This setup makes the cylindrical system's aspect ratio the reciprocal of that of the bi-planar system. We optimise magnetic field uniformity within the same spatial region as before: the central $12$~cm of the $z$-axis.

In Figs.~\ref{fig.B10_continuum_planar}a~and~\ref{fig.B11_continuum_planar}a, we present axial surface magnetisation patterns to generate uniform axial and transverse magnetic fields, respectively. The target fields along the $z$-axis produced by these patterns, calculated analytically and simulated using FEM, are shown in Figs.~\ref{fig.B10_continuum_planar}b~and~\ref{fig.B11_continuum_planar}b. The FEM simulations assume the axial surface magnetisation is uniformly distributed on circular bi-planar disks with a thickness of $z_{p0} = 1$ mm. As in the cylindrical cases, the magnetic fields generated agree closely with the target profiles and with each other. Within the target region, errors from perfect uniformity are $<1$\%, while deviations between analytic and FEM approaches are $<0.05$\%. These deviations increase when evaluating the magnetic field outside of the target region, particularly near the circular-planar magnets. They also increase in the transverse field-generating case where the surface magnetisation oscillates more greatly.

\section{Analysis}\label{ssec:analysis}
Now, we shall examine the intrinsic performance of designs developed using this methodology compared to standard equivalents and examine limitations on the realisation of these designs. Here, we shall perform this analysis for the uniform axial field-generating designs presented in section~\ref{sec:results}, although the same process could be performed for any magnet design.

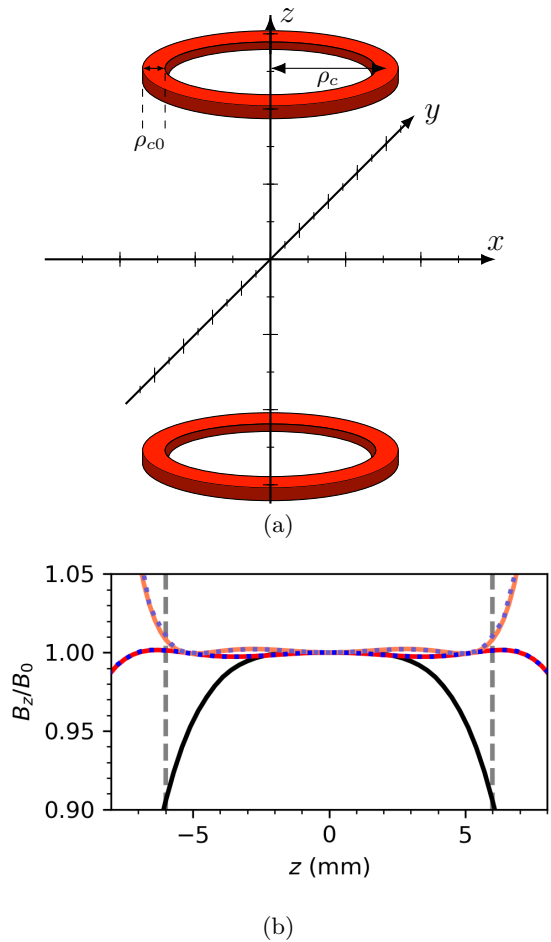
\begin{figure}[ht]
\centering
\begin{tikzpicture}[line join = bevel]

%%%%%%%%%%%%%%%%% AXES %%%%%%%%%%%%%%%%%%%%%%%%%%%%%%%%
% The axes
\pgfmathsetmacro{\zgap}{2.5}

%%%%%%%%%%%%%%%%% END OF AXES %%%%%%%%%%%%%%%%%%%%%%%%%%%%%%%%

%%%%% SHADED CYLINDER %%%%%%%%%
\pgfmathsetmacro{\cylength}{5.082/2-0.9}
\pgfmathsetmacro{\cywidth}{1.7}

% Plane
% Plane dimensions
\pgfmathsetmacro{\cylgap}{5.082}
\pgfmathsetmacro{\planethickness}{0.17}
\pgfmathsetmacro{\planetop}{\cylgap/2} % CHANGE THIS FOR SEPARATION 
\pgfmathsetmacro{\planewidth}{1.7}
\fill [lightred,opacity=1] (0,\planetop) ellipse (1.7cm and 0.5cm);
\draw[] (0,\planetop) ellipse (1.7cm and 0.5cm);

\fill [darkred,opacity=1] (-1.7,\planetop) -- (-1.7,\planetop-\planethickness) arc (180:360:1.7 and 0.5) -- (1.7,\planetop) arc (0:180:1.7 and -0.5);

\draw (-1.7,\planetop) -- (-1.7,\planetop-\planethickness);
\draw (-1.7,\planetop-\planethickness) arc (180:360:1.7 and 0.5);

\draw (1.7,\planetop-\planethickness) -- (1.7,\planetop); 

\fill [white] (0,\planetop) ellipse (1.4cm and 0.35cm);
\draw[] (0,\planetop) ellipse (1.4cm and 0.35cm);
% \draw [] (-1.35,\planetop-\planethickness) arc (180:360:1.35 and -0.35);

% \draw [blue, thick](-1.35,\planetop-\planethickness) arc (1.35:180:-1.35 and 0.35); % Bottom further
% \draw [blue, thick](-1.4,\planetop-\planethickness+0.1) arc (1.4:195:-1.4 and 0.35); % Top further
\tikzset{
    partial ellipse/.style args={#1:#2:#3}{
        insert path={+ (#1:#3) arc (#1:#2:#3)}
    }
}
\filldraw[fill=darkred, draw=black]
(0,\planetop) [partial ellipse=-15:195:1.4 and 0.35] arc (1.35:180:-1.35 and 0.35);

%%%% BOTTOM RING %%%%%%%%%
\pgfmathsetmacro{\planethickness}{0.17}
\pgfmathsetmacro{\planetop}{-\cylgap/2}
\pgfmathsetmacro{\planewidth}{1.7}
\fill [lightred,opacity=1] (0,\planetop) ellipse (1.7cm and 0.5cm);
\draw[] (0,\planetop) ellipse (1.7cm and 0.5cm);

\fill [darkred,opacity=1] (-1.7,\planetop) -- (-1.7,\planetop-\planethickness) arc (180:360:1.7 and 0.5) -- (1.7,\planetop) arc (0:180:1.7 and -0.5);

\draw (-1.7,\planetop) -- (-1.7,\planetop-\planethickness);
\draw (-1.7,\planetop-\planethickness) arc (180:360:1.7 and 0.5);

\draw (1.7,\planetop-\planethickness) -- (1.7,\planetop); 

\fill [white] (0,\planetop) ellipse (1.4cm and 0.35cm);
\draw[] (0,\planetop) ellipse (1.4cm and 0.35cm);
% \draw [] (-1.35,\planetop-\planethickness) arc (180:360:1.35 and -0.35);

% \draw [blue, thick](-1.35,\planetop-\planethickness) arc (1.35:180:-1.35 and 0.35); % Bottom further
% \draw [blue, thick](-1.4,\planetop-\planethickness+0.1) arc (1.4:195:-1.4 and 0.35); % Top further
\tikzset{
    partial ellipse/.style args={#1:#2:#3}{
        insert path={+ (#1:#3) arc (#1:#2:#3)}
    }
}
\filldraw[fill=darkred, draw=black]
(0,\planetop) [partial ellipse=-15:195:1.4 and 0.35] arc (1.35:180:-1.35 and 0.35);

% \draw [blue, ultra thick] (-1.4,\planetop-\planethickness+0.1) arc (180:360:1.4 and -0.35);
% \draw [blue, ultra thick] (-1.35,\planetop-\planethickness) arc (180:360:1.35 and -0.35);

% \draw [blue] (-1.4,-4.15) arc (180:195:1.4 and -0.35);

\pgfmathsetmacro{\zgap}{2.75}

%%%%%%%%%%%%%%%%% AXES %%%%%%%%%%%%%%%%%%%%%%%%%%%%%%%%
% The axes
\draw[-{Latex[length=2mm]}, thick] (xyz cs:x=-3) -- (xyz cs:x=3) node[above, font = \large] {$x$};
\draw[-{Latex[length=2mm]}, thick] (xyz cs:z=5) -- (xyz cs:z=-5) node[right, font = \large] {$y$};
% \draw[->] (xyz cs:y=-\zgap - 1.5) -- (xyz cs:y=\zgap + 1.5) node[right, font = \large] {$z$};

% Small - X axis
\foreach \coo in {-2.5, -2,...,2.5}
{      \draw (\coo, -0.05, 0) -- (\coo, 0.05, 0);}

% Small - Y axis
\foreach \coo in {-4.5, -4,...,4.5}
{      \draw (0,-0.05,\coo) -- (0,0.05,\coo);}

% Small - Z axis
% \foreach \coo in {-3.5,-3,...,3.5}
% {  \draw (-0.05,\coo,0) -- (0.05,\coo,0);}

% Large - Z axis
% \foreach \coo in {-3,-2,...,3}
% {  \draw (-0.1,\coo,0) -- (0.1,\coo,0);}

% Large - X axis
\foreach \coo in {-2, -1,...,2}
{       \draw (\coo, -0.1, 0) -- (\coo, 0.1, 0);}

% Large - Y axis
\foreach \coo in {-4, -3,...,4}
{       \draw (0,-0.1,\coo) -- (0,0.1,\coo);}

\draw[-, thick] (xyz cs:y=-\zgap - 0.5) -- (xyz cs:y=\zgap + 0.5);
\draw[-{Latex[length=2mm]}, thick] (xyz cs:y=2.73) -- (xyz cs:y=\zgap + 0.5) node[right, font = \large] {$z$};

% \draw[-,blue, ultra thick] (0, -\cylgap/2) -- (0,\cylgap/2);

% \draw[->] (xyz cs:y=-\zgap - 1.5) -- (xyz cs:y=\zgap + 1.5) node[right, font = \large] {$z$};

% Small - Z axis
\foreach \coo in {-3,-2.5,...,3}
{  \draw (-0.05,\coo,0) -- (0.05,\coo,0);}
% Small - Z axis
% \foreach \coo in {1.5,2,...,3.5}
% {  \draw (-0.05,\coo,0) -- (0.05,\coo,0);}

% Large - Z axis
\foreach \coo in {-3,-2,...,2}
{  \draw (-0.1,\coo,0) -- (0.1,\coo,0);}
% Large - Z axis
\foreach \coo in {3}
{  \draw (-0.1,\coo,0) -- (0.1,\coo,0);}
% Large - Z axis
% \foreach \coo in {2,3}
% {  \draw (-0.1,\coo,0) -- (0.1,\coo,0);}

%%%%%%%%%%%%%%%%% END OF AXES %%%%%%%%%%%%%%%%%%%%%%%%%%%%%%%%

\node at (-\cywidth+0.1,\cylength-0.1) {$\rho_{c0}$};
\draw[{Latex[length=1mm]}-{Latex[length=1mm]}] (-\cywidth,\cylength+0.9) -- (-\cywidth+0.3,\cylength+0.9); 
\draw[-, dashed] (-\cywidth,\cylength+0.1) -- (-\cywidth,\cylength+0.85);
\draw[-, dashed] (-\cywidth+0.3,\cylength+0.1) -- (-\cywidth+0.3,\cylength+0.85);

\node at (\cywidth/3+0.2,\cylength+0.75) {$\rho_c$};
\draw[{Latex[length=2mm]}-{Latex[length=2mm]}] (0, \cylength+0.9) -- (\cywidth-0.15, \cylength+0.9);

\end{tikzpicture} \\ \small{(a)} \\ 
\includegraphics[width=0.9\columnwidth]{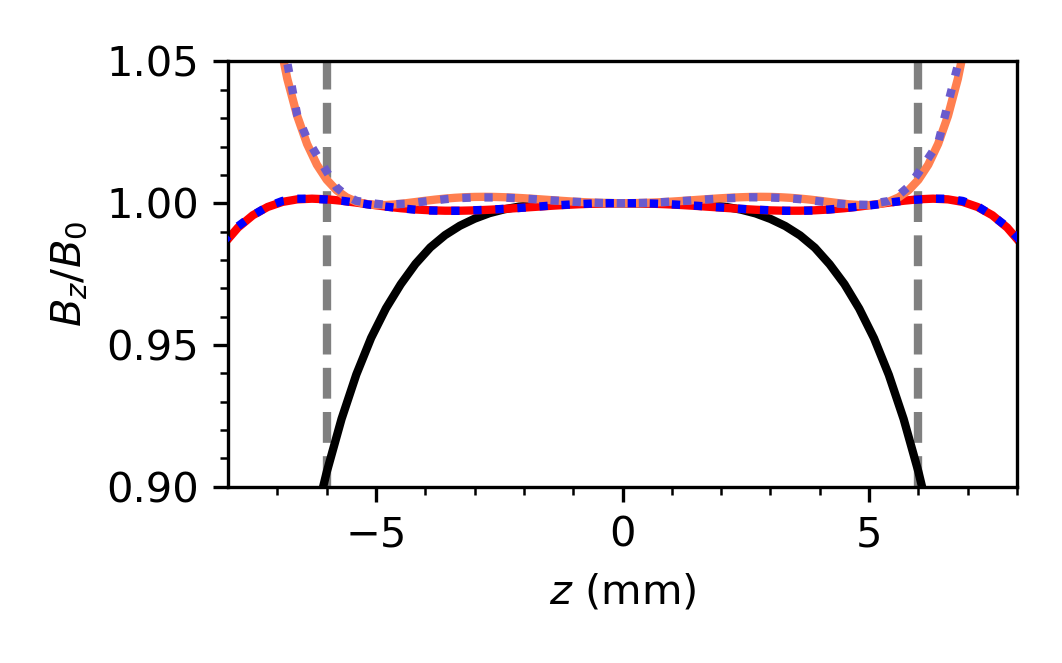} \\
\small{(b)}
\caption{(a) Schematic diagram of axially symmetric permanent magnet rings of radius $\rho_c=10$~mm and thickness $\rho_{c0}=1$~mm at axial positions $z={\pm}d_c$, where $d_c=16.94$~mm [red colouring indicates positive axial magnetisation]. (b) Normalised axial magnetic field, $B_z/B_0$, along the $z$-axis calculated for the arrangement in (a) [black solid] alongside the analytical and FEM results in Figs.~\ref{fig.B10_continuum_cylindrical}b~and~\ref{fig.B10_continuum_planar}b, respectively [darker/lighter shading for cylindrical/circular bi-planar cases].}
\label{fig.Bz_comparison_standard}
\end{figure}
In Fig.~\ref{fig.Bz_comparison_standard}a, we show a standard arrangement for generating a uniform axial magnetic field using a pair of axially magnetised loops. The loops have identical magnetisation and are positioned at axial locations $z_c = \pm1.694\rho_c$. The determination of this best separation follows Roméo and Hoult's approach in Ref.~\citep{Romeo}, whereby the uniform field is generated by nulling the leading-order error. In this case, this error is quadratic gradient of the axial field with respect to axial position, $\partial^2{B_z}/{\partial}z^2$. This gradient has a magnitude proportional to that of the $4$\textsuperscript{th} order Legendre polynomial, which has a root at $P_{4,0}\left(x=1.694\right){\approx}0$~\footnote{Another root exists at $P_{4,0}\left(x=0.362\right){\approx}0$, however, this ring magnet design generates a much smaller region with high uniformity than the more separated design.}.

In Fig.~\ref{fig.Bz_comparison_standard}b, we use FEM simulations to analyse the axial magnetic field produced by this arrangement, where the loops are three-dimensional rings with radius $\rho_c = 10$~mm and thickness $\rho_{c0} = 1$~mm. We compare these results to those generated by the surface magnetisation designs in Figs.~\ref{fig.B10_continuum_cylindrical}a~and~\ref{fig.B10_continuum_planar}a. The surface-based designs clearly provide significantly improved magnetic field uniformity compared to the standard ring magnet arrangement. Within the target region, the maximum deviation in the axial field generated by the rings is $7.61$\%, whereas for the cylindrical and circular bi-planar systems, the maximum deviations are $0.268$\% and $0.586$\%, respectively. The cylindrical system also has a smaller diameter and length than the ring pair, while the bi-planar system is much shorter in length. However, the ring system blocks access to a much smaller area and would be significantly easier to construct, and so still would be preferable in settings where these characteristics are critical.

\begin{figure}[ht]
\centering
\includegraphics[width=0.9\columnwidth]{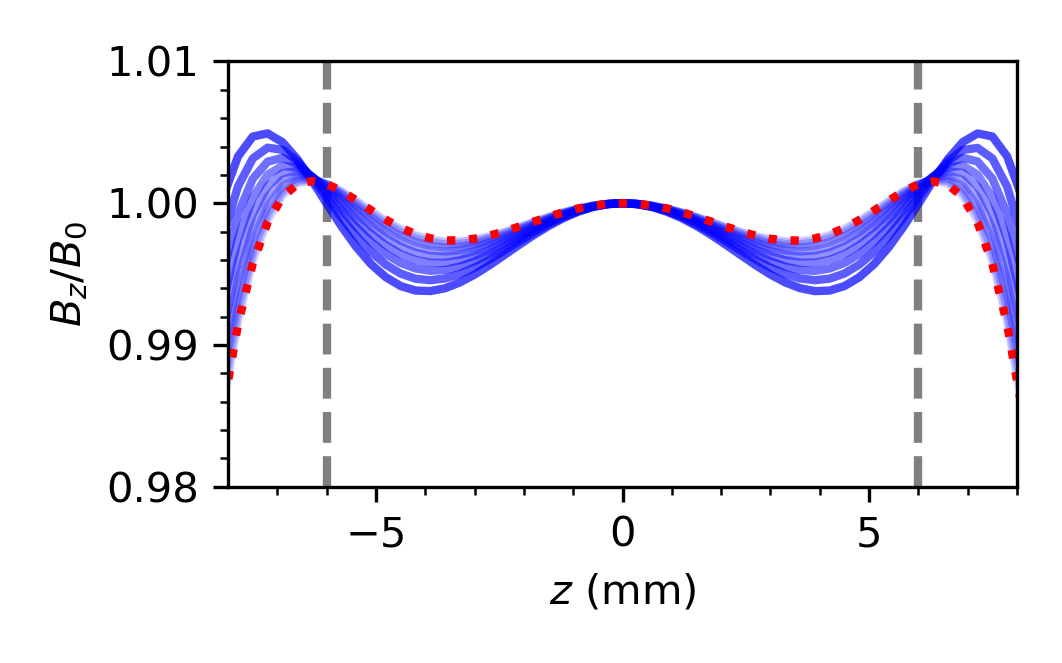} \\ \small{(a)} \\ 
\includegraphics[width=0.9\columnwidth]{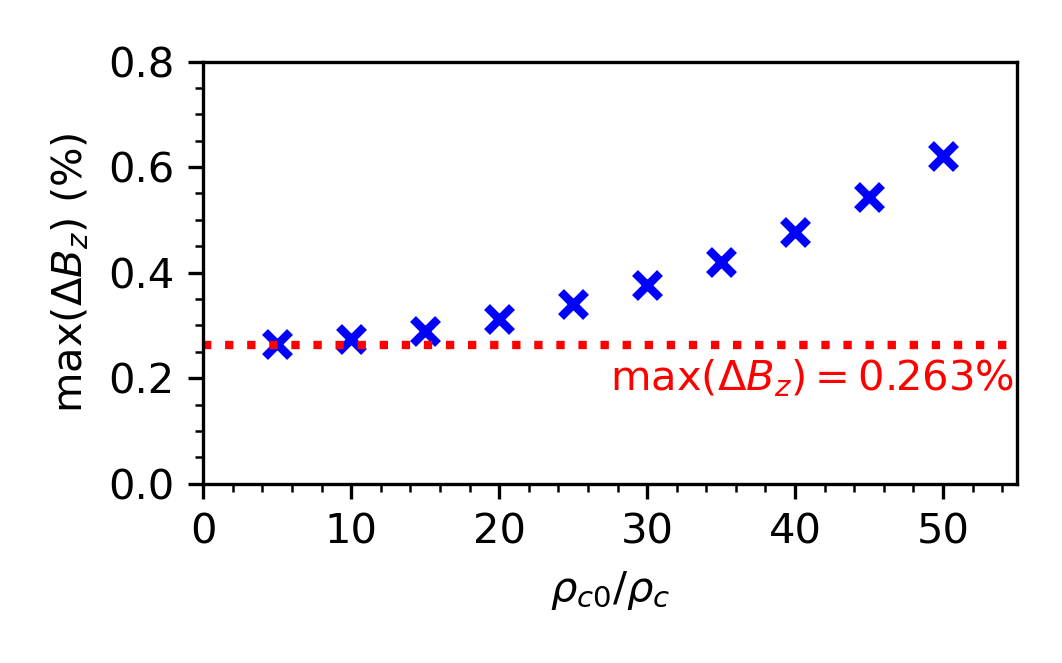} \\
\small{(b)}
\caption{(a) Normalised axial magnetic field, $B_z/B_0$, along the $z$-axis calculated from FEM simulations of the arrangement in Fig.~\ref{fig.B10_continuum_cylindrical}a of radius $\rho_c=10$~mm, where the thickness of the axially magnetised cylinder, $\rho_{c0}$, is increased from $0.5$~mm to $5$~mm in $0.5$~mm increments [light-to-dark blue shading]. The red dotted line represents the result in the thin magnet limit, where $\rho_{c0}=0.1$~mm. The edges of the target region are highlighted [dashed grey lines]. (b) Maximum axial magnetic field deviation within the target region, $\mathrm{max}({\Delta}B_z)$, as the thickness of the magnetised cylinder increases, with the dotted line representing the deviation in the thin magnet limit.}
\label{fig.Bz_comparison_thickness}
\end{figure}
Next, we use FEM simulations to assess how thick a 3D tube can be while still generating a uniform target axial field, thereby approximating the 2D cylindrical design in Fig.~\ref{fig.B10_continuum_cylindrical}a. Fig.~\ref{fig.Bz_comparison_thickness} examines the effect of varying the thickness of the volume magnetisation tube on the generated axial magnetic field and its maximum deviation within the target region. As expected, the axial field varies more as the tube's thickness increases: the maximum deviation increases from $0.263$\% at the thin magnet limit ($\rho_{c0} = 0.01\rho_c$) to $0.620$\% for the thickest tube ($\rho_{c0} = 0.5\rho_c$). Despite this increase, the error remains more than an order of magnitude smaller than that of the standard ring magnet arrangement ($7.61$\%), as shown in Fig.~\ref{fig.Bz_comparison_standard}b. This shows that even very thick 3D approximations to surface magnetisation profiles can perform better than standard designs.

While the acceptable error threshold depends on the specific application and design requirements, a general guideline is to maintain $\rho_{c0} \leq 0.1\rho_c$ for many applications. For the same total magnetisation, increasing the thickness does reduce the field magnitude slightly as more volume is located on the cylinder's larger-radius regions. However, this reduction is minor, with the axial field magnitude being only $0.96$\% lower at $\rho_{c0} = 0.50\rho_c$ compared to the thin magnet limit.

\begin{figure}[ht]
\centering
\includegraphics[width=0.9\columnwidth]{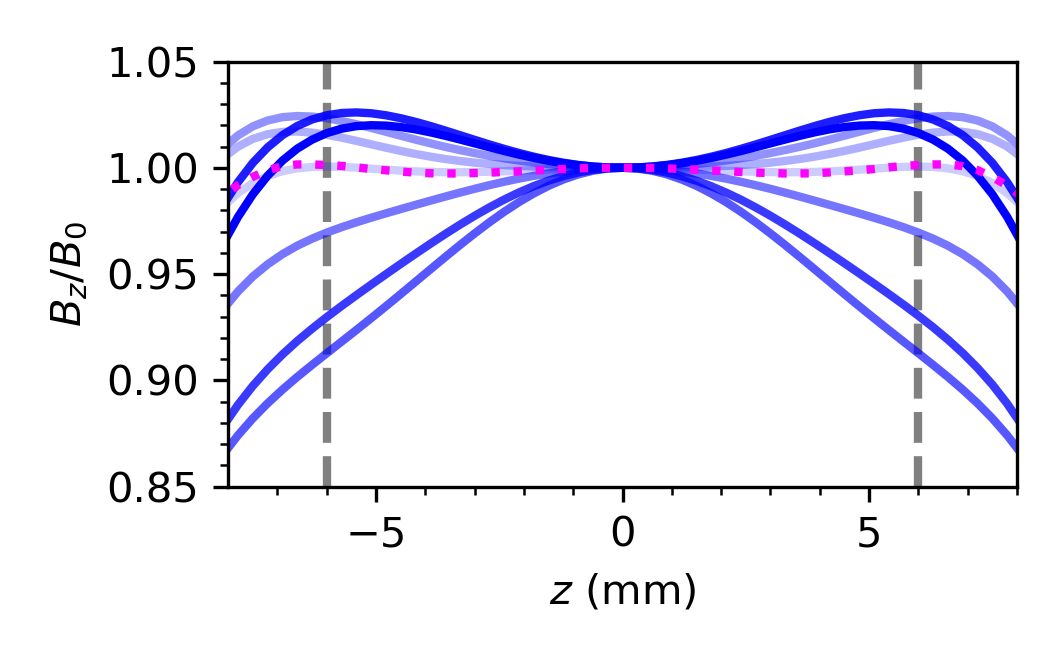} \\ \small{(a)} \\ 
\includegraphics[width=0.9\columnwidth]{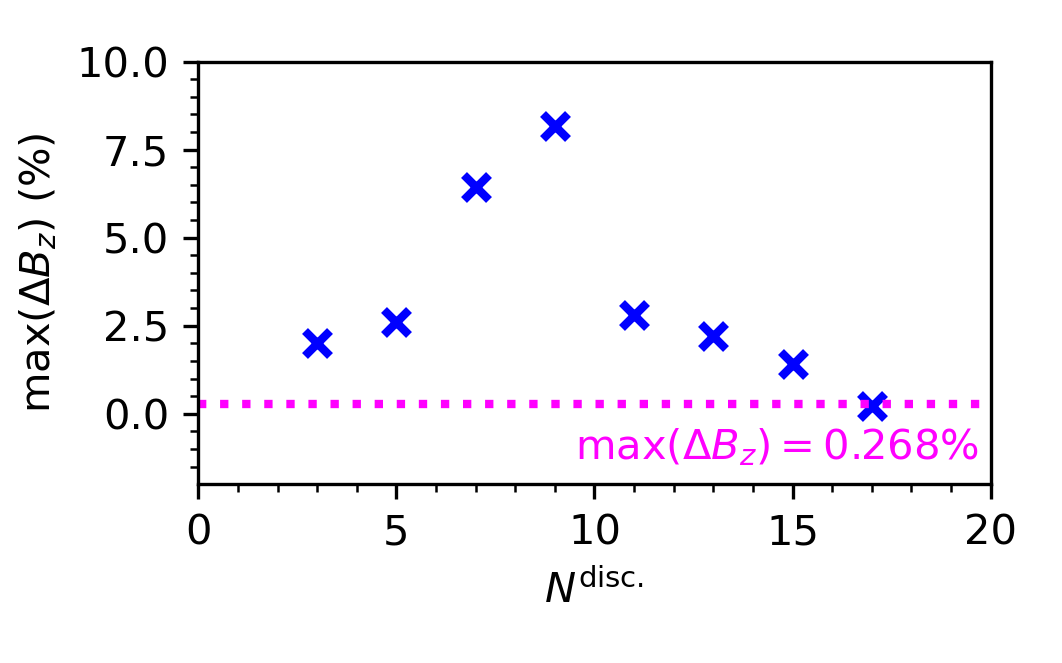} \\
\small{(b)}
\caption{(a) Normalised axial magnetic field, $B_z/B_0$, along the $z$-axis calculated from FEM simulations of the arrangement in Fig.~\ref{fig.B10_continuum_cylindrical}a of radius $\rho_c=10$~mm and fixed thickness $\rho_{c0}=1$~mm, where the number of discretisation levels, $N^\mathrm{disc.}$, is increased from $3$ to $17$ in odd intervals [dark-to-light blue shading]. The magenta dotted line represents the result in the continuum limit. The edges of the target region are highlighted [dashed grey lines]. (b) Maximum axial magnetic field deviation within the target region, $\mathrm{max}({\Delta}B_z)$, as the number of discretisation levels increases, with the dotted line representing the deviation in the continuum limit.}
\label{fig.Bz_comparison_discretisation}
\end{figure}
\begin{figure*}[!htb]
\centering
\begin{tabular}{c c c}
\includegraphics[width=0.675\columnwidth]{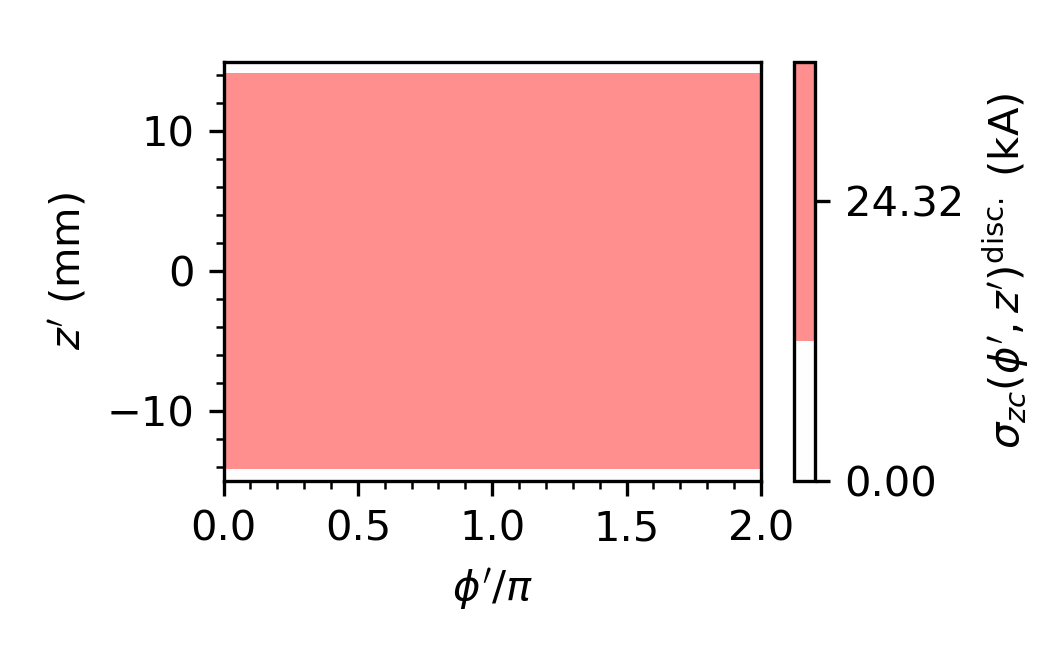} & 
\includegraphics[width=0.675\columnwidth]{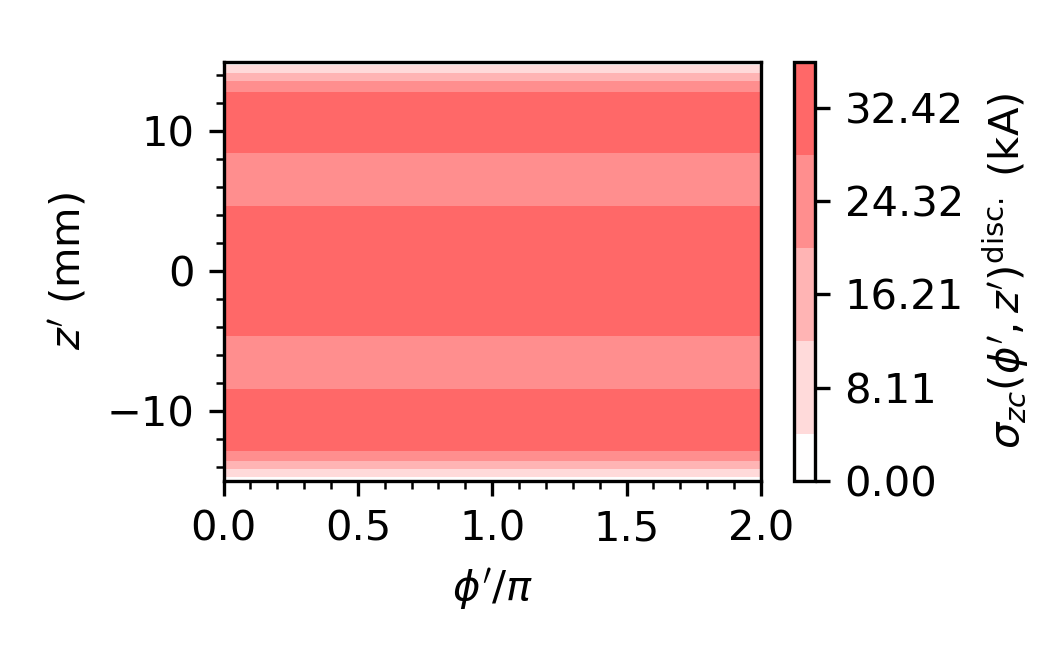} &
\includegraphics[width=0.675\columnwidth]{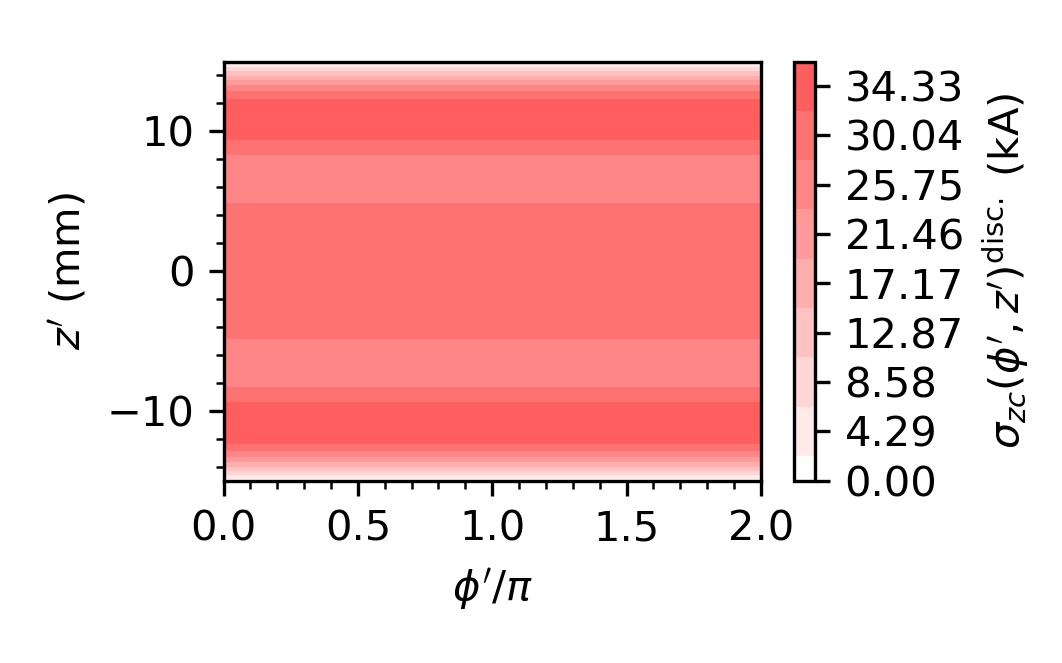} \\
\small{(a)} & \small{(b)} & \small{(c)}
\end{tabular}
\caption{Axial surface magnetisation in Fig.~\ref{fig.B10_continuum_cylindrical}a on a cylindrical surface of radius $\rho_c=10$~mm and length $L_c=30$~mm, presented at (a) $N^\mathrm{disc.}=3$, (b) $N^\mathrm{disc.}=9$, and (c) $N^\mathrm{disc.}=17$ discretisation levels. Red colouring indicates positive axial magnetisation  with increased intensity representing greater strength.}
\label{fig.contouring_discrete}
\end{figure*}
In Fig.~\ref{fig.Bz_comparison_discretisation}, we examine how varying the number of discretisation levels in the design shown in Fig.~\ref{fig.B10_continuum_cylindrical}a impacts the maximum magnetic field deviation across the target region. Examples of the discretised patterns at selected discretisation levels are presented in Fig.~\ref{fig.contouring_discrete}. The deviation displays an interesting pattern that offers valuable insight into the design of surface magnets. Initially, the error is approximately $2.5$\%, as the poorly discretised pattern resembles a uniform axial surface magnetisation across the magnet cylinder [Fig.~\ref{fig.contouring_discrete}a]. As the number of levels increases, the error rises because the spatial oscillations in the surface magnetisation pattern are inadequately represented [Fig.~\ref{fig.contouring_discrete}b]. However, when the oscillations are accurately represented with $N^\mathrm{disc.} \geq 13$, the error decreases as the discretised pattern captures more detail regarding the design. At $N^\mathrm{disc.}=19$ discretisation levels [Fig.~\ref{fig.contouring_discrete}c], the error in the discretised field across the target region drops to $0.210$\%, which is actually lower than the error intrinsic in the design in the continuum limit, where $N^\mathrm{disc.} \to \infty$. Analogous relationships are observed when approximating surface current flows as wires in coil design at a varying number of contour levels, where local minima may offer improved target field fidelity at a reduced number of contour levels~\cite{10.1016/j.jmr.2005.07.003,PhysRevApplied.14.011002}.

These analyses highlight the need for \emph{a posteriori} examination of each design for a full determination of its characteristics. If this is not possible, as a general rule, designs should allow for as many discretisation levels as possible, and the curvature regularisation weighting should be increased if the manufacturing method limits the feasible number of discretisation levels.

\section{\label{sec:conclusions}Conclusions}
In this paper, we presented a method for optimising magnetisation on a 2D surface to generate target magnetic fields. We mathematically described the relationship between axial surface magnetisation on cylindrical and circular-planar geometries and the resulting magnetic field. We used this model to design uniform axial and transverse magnetic fields using regularised least-squares optimisation. Our model closely matched numerical simulations and the uniform axial field-generating designs are over an order of magnitude more uniform across an equivalent region than the field generated by a simple pair of axially magnetised rings. Our approach assumes a fixed magnetisation and, to implement in the real-world, may require the 2D magnetisation surface to be approximated by a 3D volume. We demonstrated using FEM that an axial field-generating design on a cylinder may be approximated as being housed on a thin tube without diminishing the design performance. We find a rule-of-thumb that the tube should be about an order of magnitude thinner than its radius. Finally, we demonstrated methods to approximate the surface pattern using a discrete number of magnetisation values and then examine the resulting discrete magnetisation profile \emph{a posteriori} using FEM.

Optimised surface permanent magnets have the potential to enhance performance in various devices and disciplines, from electric machines and transducers to magnetic resonance imaging and fundamental physics. The next stage of this research is experimental validation of the designs produced using this method and then implementation into these settings. Realising the resulting magnetisation profiles experimentally, particularly those which oscillate greatly across their surfaces, may require additive manufacturing such as 3D-printing of bonded neodymium magnets~\cite{10.1016/j.scriptamat.2016.12.035} and other electromagnetic materials~\cite{TIISMUS2022102778}. Alternatively, directly manufacturing a discretised pattern using a stack of equally magnetised plates may be simpler, although a thorough analysis of both 2D-to-3D surface thickness and discretisation effects must be performed for any new design to ensure that the significant performance benefits expected from this approach are realised experimentally.

Extensive additional theoretical research may also build on this approach. Future designs may combine both cylindrical and circular-planar surface magnetisation to further optimise magnetic field fidelity. Additional theoretical work may also explore the target field design approach for radial or azimuthal surface magnetisation, potentially offering advantages in reducing internal stresses within the magnet. Magnet designs could even be encoded to account for the interaction with external magnetic shielding~\cite{PhysRevApplied.15.054004,10177829,10.1038/s41598-022-17346-1} for applications requiring strong static magnetic fields with low magnetic noise, such as in biassing ion quantum computers~\cite{Jain_2024}.

\section{Acknowledgements}\label{sec:acknowledgements}
We acknowledge support from the UK Quantum Technology Hub Sensors and Timing, funded by the Engineering and Physical Sciences Research Council (EP/M013294/1). \\

We also thank Dr Michael Packer for his enlightening contributions during discussions about magnetic field design. \\

P.J.H and M.F have a worldwide patent (WO/2021/053356) which includes some magnetic design techniques applied in this work.

\bibliography{apssamp}% Produces the bibliography via BibTeX.

\begin{thebibliography}{48}%
\makeatletter
\providecommand \@ifxundefined [1]{%
 \@ifx{#1\undefined}
}%
\providecommand \@ifnum [1]{%
 \ifnum #1\expandafter \@firstoftwo
 \else \expandafter \@secondoftwo
 \fi
}%
\providecommand \@ifx [1]{%
 \ifx #1\expandafter \@firstoftwo
 \else \expandafter \@secondoftwo
 \fi
}%
\providecommand \natexlab [1]{#1}%
\providecommand \enquote  [1]{``#1''}%
\providecommand \bibnamefont  [1]{#1}%
\providecommand \bibfnamefont [1]{#1}%
\providecommand \citenamefont [1]{#1}%
\providecommand \href@noop [0]{\@secondoftwo}%
\providecommand \href [0]{\begingroup \@sanitize@url \@href}%
\providecommand \@href[1]{\@@startlink{#1}\@@href}%
\providecommand \@@href[1]{\endgroup#1\@@endlink}%
\providecommand \@sanitize@url [0]{\catcode `\\12\catcode `\$12\catcode `\&12\catcode `\#12\catcode `\^12\catcode `\_12\catcode `\%12\relax}%
\providecommand \@@startlink[1]{}%
\providecommand \@@endlink[0]{}%
\providecommand \url  [0]{\begingroup\@sanitize@url \@url }%
\providecommand \@url [1]{\endgroup\@href {#1}{\urlprefix }}%
\providecommand \urlprefix  [0]{URL }%
\providecommand \Eprint [0]{\href }%
\providecommand \doibase [0]{https://doi.org/}%
\providecommand \selectlanguage [0]{\@gobble}%
\providecommand \bibinfo  [0]{\@secondoftwo}%
\providecommand \bibfield  [0]{\@secondoftwo}%
\providecommand \translation [1]{[#1]}%
\providecommand \BibitemOpen [0]{}%
\providecommand \bibitemStop [0]{}%
\providecommand \bibitemNoStop [0]{.\EOS\space}%
\providecommand \EOS [0]{\spacefactor3000\relax}%
\providecommand \BibitemShut  [1]{\csname bibitem#1\endcsname}%
\let\auto@bib@innerbib\@empty
%</preamble>
\bibitem [{\citenamefont {Gieras}(2010)}]{978-1-4200-6440-7}%
  \BibitemOpen
  \bibfield  {author} {\bibinfo {author} {\bibfnamefont {J.}~\bibnamefont {Gieras}},\ }\href@noop {} {\emph {\bibinfo {title} {Permanent Magnet Motor Technology: Design and Applications}}}\ (\bibinfo  {publisher} {Taylor \& Francis CRC Press Group},\ \bibinfo {year} {2010})\BibitemShut {NoStop}%
\bibitem [{\citenamefont {Zhang}\ \emph {et~al.}(2012)\citenamefont {Zhang}, \citenamefont {Xu}, \citenamefont {Junak}, \citenamefont {Fiederling}, \citenamefont {Sawczuk}, \citenamefont {Koch}, \citenamefont {Schalja}, \citenamefont {Podack},\ and\ \citenamefont {Baumgartner}}]{6425118}%
  \BibitemOpen
  \bibfield  {author} {\bibinfo {author} {\bibfnamefont {S.}~\bibnamefont {Zhang}}, \bibinfo {author} {\bibfnamefont {J.}~\bibnamefont {Xu}}, \bibinfo {author} {\bibfnamefont {J.}~\bibnamefont {Junak}}, \bibinfo {author} {\bibfnamefont {D.}~\bibnamefont {Fiederling}}, \bibinfo {author} {\bibfnamefont {G.}~\bibnamefont {Sawczuk}}, \bibinfo {author} {\bibfnamefont {M.}~\bibnamefont {Koch}}, \bibinfo {author} {\bibfnamefont {A.}~\bibnamefont {Schalja}}, \bibinfo {author} {\bibfnamefont {M.}~\bibnamefont {Podack}},\ and\ \bibinfo {author} {\bibfnamefont {J.}~\bibnamefont {Baumgartner}},\ }in\ \href {https://doi.org/10.1109/EDPC.2012.6425118} {\emph {\bibinfo {booktitle} {2012 2nd International Electric Drives Production Conference (EDPC)}}}\ (\bibinfo {year} {2012})\ pp.\ \bibinfo {pages} {1--11}\BibitemShut {NoStop}%
\bibitem [{\citenamefont {Pham}\ \emph {et~al.}(2021)\citenamefont {Pham}, \citenamefont {Kwon},\ and\ \citenamefont {Foster}}]{en14020283}%
  \BibitemOpen
  \bibfield  {author} {\bibinfo {author} {\bibfnamefont {T.}~\bibnamefont {Pham}}, \bibinfo {author} {\bibfnamefont {P.}~\bibnamefont {Kwon}},\ and\ \bibinfo {author} {\bibfnamefont {S.}~\bibnamefont {Foster}},\ }\bibfield  {journal} {\bibinfo  {journal} {Energies}\ }\textbf {\bibinfo {volume} {14}},\ \href {https://doi.org/10.3390/en14020283} {10.3390/en14020283} (\bibinfo {year} {2021})\BibitemShut {NoStop}%
\bibitem [{\citenamefont {Wang}\ \emph {et~al.}(1999)\citenamefont {Wang}, \citenamefont {Jewell},\ and\ \citenamefont {Howe}}]{764898}%
  \BibitemOpen
  \bibfield  {author} {\bibinfo {author} {\bibfnamefont {J.}~\bibnamefont {Wang}}, \bibinfo {author} {\bibfnamefont {G.}~\bibnamefont {Jewell}},\ and\ \bibinfo {author} {\bibfnamefont {D.}~\bibnamefont {Howe}},\ }\href {https://doi.org/10.1109/20.764898} {\bibfield  {journal} {\bibinfo  {journal} {IEEE Transactions on Magnetics}\ }\textbf {\bibinfo {volume} {35}},\ \bibinfo {pages} {1986} (\bibinfo {year} {1999})}\BibitemShut {NoStop}%
\bibitem [{\citenamefont {McGarry}\ \emph {et~al.}(2019)\citenamefont {McGarry}, \citenamefont {McDonald},\ and\ \citenamefont {Alotaibi}}]{8785119}%
  \BibitemOpen
  \bibfield  {author} {\bibinfo {author} {\bibfnamefont {C.}~\bibnamefont {McGarry}}, \bibinfo {author} {\bibfnamefont {A.}~\bibnamefont {McDonald}},\ and\ \bibinfo {author} {\bibfnamefont {N.}~\bibnamefont {Alotaibi}},\ }in\ \href {https://doi.org/10.1109/IEMDC.2019.8785119} {\emph {\bibinfo {booktitle} {2019 IEEE International Electric Machines \& Drives Conference (IEMDC)}}}\ (\bibinfo {year} {2019})\ pp.\ \bibinfo {pages} {656--663}\BibitemShut {NoStop}%
\bibitem [{\citenamefont {Geddes}(1994)}]{5343533A}%
  \BibitemOpen
  \bibfield  {author} {\bibinfo {author} {\bibfnamefont {E.~R.}\ \bibnamefont {Geddes}},\ }\href@noop {} {\bibinfo {title} {Transducer flux optimization}} (\bibinfo {year} {1994})\BibitemShut {NoStop}%
\bibitem [{\citenamefont {Morgan}\ \emph {et~al.}(2019)\citenamefont {Morgan}, \citenamefont {Keane}, \citenamefont {Harvey}, \citenamefont {Pittman},\ and\ \citenamefont {Coates}}]{US10469951B2}%
  \BibitemOpen
  \bibfield  {author} {\bibinfo {author} {\bibfnamefont {J.~N.}\ \bibnamefont {Morgan}}, \bibinfo {author} {\bibfnamefont {A.~J.}\ \bibnamefont {Keane}}, \bibinfo {author} {\bibfnamefont {M.~B.}\ \bibnamefont {Harvey}}, \bibinfo {author} {\bibfnamefont {C.~D.}\ \bibnamefont {Pittman}},\ and\ \bibinfo {author} {\bibfnamefont {R.~L.}\ \bibnamefont {Coates}},\ }\href@noop {} {\bibinfo {title} {Loudspeaker magnet and earphone assembly}} (\bibinfo {year} {2019})\BibitemShut {NoStop}%
\bibitem [{\citenamefont {Landreman}\ and\ \citenamefont {Zhu}(2021)}]{Landreman_2021}%
  \BibitemOpen
  \bibfield  {author} {\bibinfo {author} {\bibfnamefont {M.}~\bibnamefont {Landreman}}\ and\ \bibinfo {author} {\bibfnamefont {C.}~\bibnamefont {Zhu}},\ }\href {https://doi.org/10.1088/1361-6587/abd13d} {\bibfield  {journal} {\bibinfo  {journal} {Plasma Physics and Controlled Fusion}\ }\textbf {\bibinfo {volume} {63}},\ \bibinfo {pages} {035001} (\bibinfo {year} {2021})}\BibitemShut {NoStop}%
\bibitem [{\citenamefont {Lu}\ \emph {et~al.}(2022)\citenamefont {Lu}, \citenamefont {Xu}, \citenamefont {Chen}, \citenamefont {Zhang}, \citenamefont {Chen}, \citenamefont {Ye}, \citenamefont {Guo},\ and\ \citenamefont {Wan}}]{LU2022100709}%
  \BibitemOpen
  \bibfield  {author} {\bibinfo {author} {\bibfnamefont {Z.}~\bibnamefont {Lu}}, \bibinfo {author} {\bibfnamefont {G.}~\bibnamefont {Xu}}, \bibinfo {author} {\bibfnamefont {D.}~\bibnamefont {Chen}}, \bibinfo {author} {\bibfnamefont {X.}~\bibnamefont {Zhang}}, \bibinfo {author} {\bibfnamefont {L.}~\bibnamefont {Chen}}, \bibinfo {author} {\bibfnamefont {M.}~\bibnamefont {Ye}}, \bibinfo {author} {\bibfnamefont {H.}~\bibnamefont {Guo}},\ and\ \bibinfo {author} {\bibfnamefont {B.}~\bibnamefont {Wan}},\ }\href {https://doi.org/https://doi.org/10.1016/j.xcrp.2021.100709} {\bibfield  {journal} {\bibinfo  {journal} {Cell Reports Physical Science}\ }\textbf {\bibinfo {volume} {3}},\ \bibinfo {pages} {100709} (\bibinfo {year} {2022})}\BibitemShut {NoStop}%
\bibitem [{\citenamefont {Qian}\ \emph {et~al.}(2022)\citenamefont {Qian}, \citenamefont {Zarnstorff}, \citenamefont {Bishop}, \citenamefont {Chamblis}, \citenamefont {Dominguez}, \citenamefont {Pagano}, \citenamefont {Patch},\ and\ \citenamefont {Zhu}}]{Qian_2022}%
  \BibitemOpen
  \bibfield  {author} {\bibinfo {author} {\bibfnamefont {T.}~\bibnamefont {Qian}}, \bibinfo {author} {\bibfnamefont {M.}~\bibnamefont {Zarnstorff}}, \bibinfo {author} {\bibfnamefont {D.}~\bibnamefont {Bishop}}, \bibinfo {author} {\bibfnamefont {A.}~\bibnamefont {Chamblis}}, \bibinfo {author} {\bibfnamefont {A.}~\bibnamefont {Dominguez}}, \bibinfo {author} {\bibfnamefont {C.}~\bibnamefont {Pagano}}, \bibinfo {author} {\bibfnamefont {D.}~\bibnamefont {Patch}},\ and\ \bibinfo {author} {\bibfnamefont {C.}~\bibnamefont {Zhu}},\ }\href {https://doi.org/10.1088/1741-4326/ac6c99} {\bibfield  {journal} {\bibinfo  {journal} {Nuclear Fusion}\ }\textbf {\bibinfo {volume} {62}},\ \bibinfo {pages} {084001} (\bibinfo {year} {2022})}\BibitemShut {NoStop}%
\bibitem [{\citenamefont {Xu}\ \emph {et~al.}(2021)\citenamefont {Xu}, \citenamefont {Lu}, \citenamefont {Chen}, \citenamefont {Chen}, \citenamefont {Zhang}, \citenamefont {Wu}, \citenamefont {Ye},\ and\ \citenamefont {Wan}}]{Xu_2021}%
  \BibitemOpen
  \bibfield  {author} {\bibinfo {author} {\bibfnamefont {G.}~\bibnamefont {Xu}}, \bibinfo {author} {\bibfnamefont {Z.}~\bibnamefont {Lu}}, \bibinfo {author} {\bibfnamefont {D.}~\bibnamefont {Chen}}, \bibinfo {author} {\bibfnamefont {L.}~\bibnamefont {Chen}}, \bibinfo {author} {\bibfnamefont {X.}~\bibnamefont {Zhang}}, \bibinfo {author} {\bibfnamefont {X.}~\bibnamefont {Wu}}, \bibinfo {author} {\bibfnamefont {M.}~\bibnamefont {Ye}},\ and\ \bibinfo {author} {\bibfnamefont {B.}~\bibnamefont {Wan}},\ }\href {https://doi.org/10.1088/1741-4326/abcdb6} {\bibfield  {journal} {\bibinfo  {journal} {Nuclear Fusion}\ }\textbf {\bibinfo {volume} {61}},\ \bibinfo {pages} {026025} (\bibinfo {year} {2021})}\BibitemShut {NoStop}%
\bibitem [{\citenamefont {Lu}\ \emph {et~al.}(2021)\citenamefont {Lu}, \citenamefont {Xu}, \citenamefont {Chen}, \citenamefont {Chen}, \citenamefont {Zhang}, \citenamefont {Ye},\ and\ \citenamefont {Wan}}]{Lu_2021}%
  \BibitemOpen
  \bibfield  {author} {\bibinfo {author} {\bibfnamefont {Z.}~\bibnamefont {Lu}}, \bibinfo {author} {\bibfnamefont {G.}~\bibnamefont {Xu}}, \bibinfo {author} {\bibfnamefont {D.}~\bibnamefont {Chen}}, \bibinfo {author} {\bibfnamefont {L.}~\bibnamefont {Chen}}, \bibinfo {author} {\bibfnamefont {X.}~\bibnamefont {Zhang}}, \bibinfo {author} {\bibfnamefont {M.}~\bibnamefont {Ye}},\ and\ \bibinfo {author} {\bibfnamefont {B.}~\bibnamefont {Wan}},\ }\href {https://doi.org/10.1088/1741-4326/ac1710} {\bibfield  {journal} {\bibinfo  {journal} {Nuclear Fusion}\ }\textbf {\bibinfo {volume} {61}},\ \bibinfo {pages} {106028} (\bibinfo {year} {2021})}\BibitemShut {NoStop}%
\bibitem [{\citenamefont {Paulsen}\ \emph {et~al.}(2008)\citenamefont {Paulsen}, \citenamefont {Franck}, \citenamefont {Demas},\ and\ \citenamefont {Bouchard}}]{4711303}%
  \BibitemOpen
  \bibfield  {author} {\bibinfo {author} {\bibfnamefont {J.~L.}\ \bibnamefont {Paulsen}}, \bibinfo {author} {\bibfnamefont {J.}~\bibnamefont {Franck}}, \bibinfo {author} {\bibfnamefont {V.}~\bibnamefont {Demas}},\ and\ \bibinfo {author} {\bibfnamefont {L.-S.}\ \bibnamefont {Bouchard}},\ }\href {https://doi.org/10.1109/TMAG.2008.2001697} {\bibfield  {journal} {\bibinfo  {journal} {IEEE Transactions on Magnetics}\ }\textbf {\bibinfo {volume} {44}},\ \bibinfo {pages} {4582} (\bibinfo {year} {2008})}\BibitemShut {NoStop}%
\bibitem [{\citenamefont {Florio}\ \emph {et~al.}(2018)\citenamefont {Florio}, \citenamefont {Sinha},\ and\ \citenamefont {Sundararaman}}]{8302405}%
  \BibitemOpen
  \bibfield  {author} {\bibinfo {author} {\bibfnamefont {F.}~\bibnamefont {Florio}}, \bibinfo {author} {\bibfnamefont {G.}~\bibnamefont {Sinha}},\ and\ \bibinfo {author} {\bibfnamefont {R.}~\bibnamefont {Sundararaman}},\ }\href {https://doi.org/10.1109/TMAG.2018.2795592} {\bibfield  {journal} {\bibinfo  {journal} {IEEE Transactions on Magnetics}\ }\textbf {\bibinfo {volume} {54}},\ \bibinfo {pages} {1} (\bibinfo {year} {2018})}\BibitemShut {NoStop}%
\bibitem [{\citenamefont {Parsagian}\ and\ \citenamefont {Kleinert}(2015)}]{10.1119/1.4930080}%
  \BibitemOpen
  \bibfield  {author} {\bibinfo {author} {\bibfnamefont {A.}~\bibnamefont {Parsagian}}\ and\ \bibinfo {author} {\bibfnamefont {M.}~\bibnamefont {Kleinert}},\ }\href {https://doi.org/10.1119/1.4930080} {\bibfield  {journal} {\bibinfo  {journal} {American Journal of Physics}\ }\textbf {\bibinfo {volume} {83}},\ \bibinfo {pages} {892} (\bibinfo {year} {2015})}\BibitemShut {NoStop}%
\bibitem [{\citenamefont {Cheiney}\ \emph {et~al.}(2011)\citenamefont {Cheiney}, \citenamefont {Carraz}, \citenamefont {Bartoszek-Bober}, \citenamefont {Faure}, \citenamefont {Vermersch}, \citenamefont {Fabre}, \citenamefont {Gattobigio}, \citenamefont {Lahaye}, \citenamefont {guéry odelin},\ and\ \citenamefont {Mathevet}}]{10.1063/1.3600897}%
  \BibitemOpen
  \bibfield  {author} {\bibinfo {author} {\bibfnamefont {P.}~\bibnamefont {Cheiney}}, \bibinfo {author} {\bibfnamefont {O.}~\bibnamefont {Carraz}}, \bibinfo {author} {\bibfnamefont {D.}~\bibnamefont {Bartoszek-Bober}}, \bibinfo {author} {\bibfnamefont {S.}~\bibnamefont {Faure}}, \bibinfo {author} {\bibfnamefont {F.}~\bibnamefont {Vermersch}}, \bibinfo {author} {\bibfnamefont {C.}~\bibnamefont {Fabre}}, \bibinfo {author} {\bibfnamefont {G.~L.}\ \bibnamefont {Gattobigio}}, \bibinfo {author} {\bibfnamefont {T.}~\bibnamefont {Lahaye}}, \bibinfo {author} {\bibfnamefont {D.}~\bibnamefont {guéry odelin}},\ and\ \bibinfo {author} {\bibfnamefont {R.}~\bibnamefont {Mathevet}},\ }\href {https://doi.org/10.1063/1.3600897} {\bibfield  {journal} {\bibinfo  {journal} {The Review of scientific instruments}\ }\textbf {\bibinfo {volume} {82}},\ \bibinfo {pages} {063115} (\bibinfo {year} {2011})}\BibitemShut {NoStop}%
\bibitem [{\citenamefont {Isoardi}\ \emph {et~al.}(2023)\citenamefont {Isoardi}, \citenamefont {Ferretti}, \citenamefont {Bonmassar},\ and\ \citenamefont {Manassero}}]{ISOARDI2023111792}%
  \BibitemOpen
  \bibfield  {author} {\bibinfo {author} {\bibfnamefont {T.}~\bibnamefont {Isoardi}}, \bibinfo {author} {\bibfnamefont {A.}~\bibnamefont {Ferretti}}, \bibinfo {author} {\bibfnamefont {L.}~\bibnamefont {Bonmassar}},\ and\ \bibinfo {author} {\bibfnamefont {P.}~\bibnamefont {Manassero}},\ }\href {https://doi.org/https://doi.org/10.1016/j.vacuum.2022.111792} {\bibfield  {journal} {\bibinfo  {journal} {Vacuum}\ }\textbf {\bibinfo {volume} {209}},\ \bibinfo {pages} {111792} (\bibinfo {year} {2023})}\BibitemShut {NoStop}%
\bibitem [{\citenamefont {Lee}\ and\ \citenamefont {Kim}(2018)}]{LEE2018869}%
  \BibitemOpen
  \bibfield  {author} {\bibinfo {author} {\bibfnamefont {G.~H.}\ \bibnamefont {Lee}}\ and\ \bibinfo {author} {\bibfnamefont {H.~R.}\ \bibnamefont {Kim}},\ }\href {https://doi.org/https://doi.org/10.1016/j.net.2018.04.010} {\bibfield  {journal} {\bibinfo  {journal} {Nuclear Engineering and Technology}\ }\textbf {\bibinfo {volume} {50}},\ \bibinfo {pages} {869} (\bibinfo {year} {2018})}\BibitemShut {NoStop}%
\bibitem [{\citenamefont {Madkhaly}\ \emph {et~al.}(2021)\citenamefont {Madkhaly}, \citenamefont {Coles}, \citenamefont {Morley}, \citenamefont {Colquhoun}, \citenamefont {Fromhold}, \citenamefont {Cooper},\ and\ \citenamefont {Hackerm\"uller}}]{PRXQuantum.2.030326}%
  \BibitemOpen
  \bibfield  {author} {\bibinfo {author} {\bibfnamefont {S.}~\bibnamefont {Madkhaly}}, \bibinfo {author} {\bibfnamefont {L.}~\bibnamefont {Coles}}, \bibinfo {author} {\bibfnamefont {C.}~\bibnamefont {Morley}}, \bibinfo {author} {\bibfnamefont {C.}~\bibnamefont {Colquhoun}}, \bibinfo {author} {\bibfnamefont {T.}~\bibnamefont {Fromhold}}, \bibinfo {author} {\bibfnamefont {N.}~\bibnamefont {Cooper}},\ and\ \bibinfo {author} {\bibfnamefont {L.}~\bibnamefont {Hackerm\"uller}},\ }\href {https://doi.org/10.1103/PRXQuantum.2.030326} {\bibfield  {journal} {\bibinfo  {journal} {PRX Quantum}\ }\textbf {\bibinfo {volume} {2}},\ \bibinfo {pages} {030326} (\bibinfo {year} {2021})}\BibitemShut {NoStop}%
\bibitem [{\citenamefont {Furlani}(1997)}]{EPFurlani_1997}%
  \BibitemOpen
  \bibfield  {author} {\bibinfo {author} {\bibfnamefont {E.~P.}\ \bibnamefont {Furlani}},\ }\href {https://doi.org/10.1088/0022-3727/30/13/004} {\bibfield  {journal} {\bibinfo  {journal} {Journal of Physics D: Applied Physics}\ }\textbf {\bibinfo {volume} {30}},\ \bibinfo {pages} {1846} (\bibinfo {year} {1997})}\BibitemShut {NoStop}%
\bibitem [{\citenamefont {Liu}\ \emph {et~al.}(2021)\citenamefont {Liu}, \citenamefont {Wu}, \citenamefont {Zhang}, \citenamefont {Liu}, \citenamefont {Huang}, \citenamefont {Zhou}, \citenamefont {Jia}, \citenamefont {Zhai},\ and\ \citenamefont {Sun}}]{LIU2021165428}%
  \BibitemOpen
  \bibfield  {author} {\bibinfo {author} {\bibfnamefont {Y.}~\bibnamefont {Liu}}, \bibinfo {author} {\bibfnamefont {Q.}~\bibnamefont {Wu}}, \bibinfo {author} {\bibfnamefont {X.}~\bibnamefont {Zhang}}, \bibinfo {author} {\bibfnamefont {J.}~\bibnamefont {Liu}}, \bibinfo {author} {\bibfnamefont {W.}~\bibnamefont {Huang}}, \bibinfo {author} {\bibfnamefont {Y.}~\bibnamefont {Zhou}}, \bibinfo {author} {\bibfnamefont {Z.}~\bibnamefont {Jia}}, \bibinfo {author} {\bibfnamefont {Y.}~\bibnamefont {Zhai}},\ and\ \bibinfo {author} {\bibfnamefont {L.}~\bibnamefont {Sun}},\ }\href {https://doi.org/https://doi.org/10.1016/j.nima.2021.165428} {\bibfield  {journal} {\bibinfo  {journal} {Nuclear Instruments and Methods in Physics Research Section A: Accelerators, Spectrometers, Detectors and Associated Equipment}\ }\textbf {\bibinfo {volume} {1006}},\ \bibinfo {pages} {165428} (\bibinfo {year} {2021})}\BibitemShut {NoStop}%
\bibitem [{\citenamefont {Roméo}\ and\ \citenamefont {Hoult}(1984)}]{Romeo}%
  \BibitemOpen
  \bibfield  {author} {\bibinfo {author} {\bibfnamefont {F.}~\bibnamefont {Roméo}}\ and\ \bibinfo {author} {\bibfnamefont {D.~I.}\ \bibnamefont {Hoult}},\ }\href {https://doi.org/https://doi.org/10.1002/mrm.1910010107} {\bibfield  {journal} {\bibinfo  {journal} {Magnetic Resonance in Medicine}\ }\textbf {\bibinfo {volume} {1}},\ \bibinfo {pages} {44} (\bibinfo {year} {1984})}\BibitemShut {NoStop}%
\bibitem [{\citenamefont {Peng}\ \emph {et~al.}(2004)\citenamefont {Peng}, \citenamefont {McMurry},\ and\ \citenamefont {Coey}}]{PENG2004165}%
  \BibitemOpen
  \bibfield  {author} {\bibinfo {author} {\bibfnamefont {Q.}~\bibnamefont {Peng}}, \bibinfo {author} {\bibfnamefont {S.}~\bibnamefont {McMurry}},\ and\ \bibinfo {author} {\bibfnamefont {J.}~\bibnamefont {Coey}},\ }\href {https://doi.org/https://doi.org/10.1016/S0304-8853(03)00494-3} {\bibfield  {journal} {\bibinfo  {journal} {Journal of Magnetism and Magnetic Materials}\ }\textbf {\bibinfo {volume} {268}},\ \bibinfo {pages} {165} (\bibinfo {year} {2004})}\BibitemShut {NoStop}%
\bibitem [{\citenamefont {Selvaggi}\ \emph {et~al.}(2004)\citenamefont {Selvaggi}, \citenamefont {Salon}, \citenamefont {Kwon},\ and\ \citenamefont {Chari}}]{10.1109/TMAG.2004.831653}%
  \BibitemOpen
  \bibfield  {author} {\bibinfo {author} {\bibfnamefont {J.}~\bibnamefont {Selvaggi}}, \bibinfo {author} {\bibfnamefont {S.}~\bibnamefont {Salon}}, \bibinfo {author} {\bibfnamefont {O.-M.}\ \bibnamefont {Kwon}},\ and\ \bibinfo {author} {\bibfnamefont {M.}~\bibnamefont {Chari}},\ }\href {https://doi.org/10.1109/TMAG.2004.831653} {\bibfield  {journal} {\bibinfo  {journal} {IEEE Transactions on Magnetics}\ }\textbf {\bibinfo {volume} {40}},\ \bibinfo {pages} {3278} (\bibinfo {year} {2004})}\BibitemShut {NoStop}%
\bibitem [{\citenamefont {Schmied}\ \emph {et~al.}(2010)\citenamefont {Schmied}, \citenamefont {Leibfried}, \citenamefont {Spreeuw},\ and\ \citenamefont {Whitlock}}]{Schmied_2010}%
  \BibitemOpen
  \bibfield  {author} {\bibinfo {author} {\bibfnamefont {R.}~\bibnamefont {Schmied}}, \bibinfo {author} {\bibfnamefont {D.}~\bibnamefont {Leibfried}}, \bibinfo {author} {\bibfnamefont {R.~J.~C.}\ \bibnamefont {Spreeuw}},\ and\ \bibinfo {author} {\bibfnamefont {S.}~\bibnamefont {Whitlock}},\ }\href {https://doi.org/10.1088/1367-2630/12/10/103029} {\bibfield  {journal} {\bibinfo  {journal} {New Journal of Physics}\ }\textbf {\bibinfo {volume} {12}},\ \bibinfo {pages} {103029} (\bibinfo {year} {2010})}\BibitemShut {NoStop}%
\bibitem [{\citenamefont {Noguchi}\ \emph {et~al.}(2014)\citenamefont {Noguchi}, \citenamefont {Kim}, \citenamefont {Hahn},\ and\ \citenamefont {Iwasa}}]{6749021}%
  \BibitemOpen
  \bibfield  {author} {\bibinfo {author} {\bibfnamefont {S.}~\bibnamefont {Noguchi}}, \bibinfo {author} {\bibfnamefont {S.}~\bibnamefont {Kim}}, \bibinfo {author} {\bibfnamefont {S.}~\bibnamefont {Hahn}},\ and\ \bibinfo {author} {\bibfnamefont {Y.}~\bibnamefont {Iwasa}},\ }\href {https://doi.org/10.1109/TMAG.2013.2276736} {\bibfield  {journal} {\bibinfo  {journal} {IEEE Transactions on Magnetics}\ }\textbf {\bibinfo {volume} {50}},\ \bibinfo {pages} {605} (\bibinfo {year} {2014})}\BibitemShut {NoStop}%
\bibitem [{\citenamefont {{Turner}}(1986)}]{10.1088/0022-3727/19/8/001}%
  \BibitemOpen
  \bibfield  {author} {\bibinfo {author} {\bibfnamefont {R.}~\bibnamefont {{Turner}}},\ }\href {https://doi.org/10.1088/0022-3727/19/8/001} {\bibfield  {journal} {\bibinfo  {journal} {Journal of Physics D Applied Physics}\ }\textbf {\bibinfo {volume} {19}},\ \bibinfo {pages} {L147} (\bibinfo {year} {1986})}\BibitemShut {NoStop}%
\bibitem [{\citenamefont {Forbes}\ and\ \citenamefont {Crozier}(2001)}]{10.1088/0022-3727/34/24/305}%
  \BibitemOpen
  \bibfield  {author} {\bibinfo {author} {\bibfnamefont {L.~K.}\ \bibnamefont {Forbes}}\ and\ \bibinfo {author} {\bibfnamefont {S.}~\bibnamefont {Crozier}},\ }\href {https://doi.org/10.1088/0022-3727/34/24/305} {\bibfield  {journal} {\bibinfo  {journal} {Journal of Physics D: Applied Physics}\ }\textbf {\bibinfo {volume} {34}},\ \bibinfo {pages} {3447} (\bibinfo {year} {2001})}\BibitemShut {NoStop}%
\bibitem [{\citenamefont {Forbes}\ and\ \citenamefont {Crozier}(2002{\natexlab{a}})}]{10.1088/0022-3727/35/9/303}%
  \BibitemOpen
  \bibfield  {author} {\bibinfo {author} {\bibfnamefont {L.~K.}\ \bibnamefont {Forbes}}\ and\ \bibinfo {author} {\bibfnamefont {S.}~\bibnamefont {Crozier}},\ }\href {https://doi.org/10.1088/0022-3727/35/9/303} {\bibfield  {journal} {\bibinfo  {journal} {Journal of Physics D: Applied Physics}\ }\textbf {\bibinfo {volume} {35}},\ \bibinfo {pages} {839} (\bibinfo {year} {2002}{\natexlab{a}})}\BibitemShut {NoStop}%
\bibitem [{\citenamefont {Forbes}\ and\ \citenamefont {Crozier}(2002{\natexlab{b}})}]{10.1088/0022-3727/36/2/302}%
  \BibitemOpen
  \bibfield  {author} {\bibinfo {author} {\bibfnamefont {L.~K.}\ \bibnamefont {Forbes}}\ and\ \bibinfo {author} {\bibfnamefont {S.}~\bibnamefont {Crozier}},\ }\href {https://doi.org/10.1088/0022-3727/36/2/302} {\bibfield  {journal} {\bibinfo  {journal} {Journal of Physics D: Applied Physics}\ }\textbf {\bibinfo {volume} {36}},\ \bibinfo {pages} {68} (\bibinfo {year} {2002}{\natexlab{b}})}\BibitemShut {NoStop}%
\bibitem [{\citenamefont {Carlson}\ \emph {et~al.}(1990)\citenamefont {Carlson}, \citenamefont {Derby}, \citenamefont {Hawryszko},\ and\ \citenamefont {Weideman}}]{693569}%
  \BibitemOpen
  \bibfield  {author} {\bibinfo {author} {\bibfnamefont {J.}~\bibnamefont {Carlson}}, \bibinfo {author} {\bibfnamefont {K.}~\bibnamefont {Derby}}, \bibinfo {author} {\bibfnamefont {C.}~\bibnamefont {Hawryszko}},\ and\ \bibinfo {author} {\bibfnamefont {M.}~\bibnamefont {Weideman}},\ }in\ \href {https://doi.org/10.1109/NSSMIC.1990.693569} {\emph {\bibinfo {booktitle} {1990 IEEE Nuclear Science Symposium Conference Record}}}\ (\bibinfo {year} {1990})\ pp.\ \bibinfo {pages} {1386--1390}\BibitemShut {NoStop}%
\bibitem [{\citenamefont {Jackson}(1962)}]{JacksonEMscalar}%
  \BibitemOpen
  \bibfield  {author} {\bibinfo {author} {\bibfnamefont {J.~D.}\ \bibnamefont {Jackson}},\ }\bibinfo {title} {Classical electrodynamics}\ (\bibinfo  {publisher} {Wiley},\ \bibinfo {address} {New York, {NY}},\ \bibinfo {year} {1962})\ pp.\ \bibinfo {pages} {156--158},\ \bibinfo {edition} {1st}\ ed.\BibitemShut {Stop}%
\bibitem [{\citenamefont {Cabada}(2014)}]{10.1007/978-1-4614-9506-2}%
  \BibitemOpen
  \bibfield  {author} {\bibinfo {author} {\bibfnamefont {A.}~\bibnamefont {Cabada}},\ }\href {https://doi.org/10.1007/978-1-4614-9506-2} {\emph {\bibinfo {title} {Green’s Functions in the Theory of Ordinary Differential Equations}}}\ (\bibinfo  {publisher} {Springer Briefs in Mathematics},\ \bibinfo {year} {2014})\BibitemShut {NoStop}%
\bibitem [{\citenamefont {Mäkinen}\ \emph {et~al.}(2020)\citenamefont {Mäkinen}, \citenamefont {Zetter}, \citenamefont {Iivanainen}, \citenamefont {Zevenhoven}, \citenamefont {Parkkonen},\ and\ \citenamefont {Ilmoniemi}}]{mkinen2020magneticfield}%
  \BibitemOpen
  \bibfield  {author} {\bibinfo {author} {\bibfnamefont {A.~J.}\ \bibnamefont {Mäkinen}}, \bibinfo {author} {\bibfnamefont {R.}~\bibnamefont {Zetter}}, \bibinfo {author} {\bibfnamefont {J.}~\bibnamefont {Iivanainen}}, \bibinfo {author} {\bibfnamefont {K.~C.~J.}\ \bibnamefont {Zevenhoven}}, \bibinfo {author} {\bibfnamefont {L.}~\bibnamefont {Parkkonen}},\ and\ \bibinfo {author} {\bibfnamefont {R.~J.}\ \bibnamefont {Ilmoniemi}},\ }\href {https://doi.org/10.1063/5.0016090} {\bibfield  {journal} {\bibinfo  {journal} {Journal of Applied Physics}\ }\textbf {\bibinfo {volume} {128}},\ \bibinfo {pages} {063906} (\bibinfo {year} {2020})}\BibitemShut {NoStop}%
\bibitem [{\citenamefont {Zetter}\ \emph {et~al.}(2020)\citenamefont {Zetter}, \citenamefont {J.~Mäkinen}, \citenamefont {Iivanainen}, \citenamefont {Zevenhoven}, \citenamefont {Ilmoniemi},\ and\ \citenamefont {Parkkonen}}]{zetter2020magneticfield}%
  \BibitemOpen
  \bibfield  {author} {\bibinfo {author} {\bibfnamefont {R.}~\bibnamefont {Zetter}}, \bibinfo {author} {\bibfnamefont {A.}~\bibnamefont {J.~Mäkinen}}, \bibinfo {author} {\bibfnamefont {J.}~\bibnamefont {Iivanainen}}, \bibinfo {author} {\bibfnamefont {K.~C.~J.}\ \bibnamefont {Zevenhoven}}, \bibinfo {author} {\bibfnamefont {R.~J.}\ \bibnamefont {Ilmoniemi}},\ and\ \bibinfo {author} {\bibfnamefont {L.}~\bibnamefont {Parkkonen}},\ }\href@noop {} {\bibfield  {journal} {\bibinfo  {journal} {Journal of Applied Physics}\ }\textbf {\bibinfo {volume} {128}},\ \bibinfo {pages} {063905} (\bibinfo {year} {2020})}\BibitemShut {NoStop}%
\bibitem [{\citenamefont {Wolberg}(2006)}]{10.1007/3-540-31720-1}%
  \BibitemOpen
  \bibfield  {author} {\bibinfo {author} {\bibfnamefont {J.}~\bibnamefont {Wolberg}},\ }\href {https://doi.org/10.1007/3-540-31720-1} {\emph {\bibinfo {title} {Data Analysis Using the Method of Least Squares: Extracting the Most Information from Experiments}}}\ (\bibinfo  {publisher} {Springer Science \& Business Media},\ \bibinfo {year} {2006})\ pp.\ \bibinfo {pages} {1--250}\BibitemShut {NoStop}%
\bibitem [{\citenamefont {Lopez}\ \emph {et~al.}(2010)\citenamefont {Lopez}, \citenamefont {Poole},\ and\ \citenamefont {Crozier}}]{10.1016/j.jmr.2010.09.002}%
  \BibitemOpen
  \bibfield  {author} {\bibinfo {author} {\bibfnamefont {H.}~\bibnamefont {Lopez}}, \bibinfo {author} {\bibfnamefont {M.}~\bibnamefont {Poole}},\ and\ \bibinfo {author} {\bibfnamefont {S.}~\bibnamefont {Crozier}},\ }\href {https://doi.org/10.1016/j.jmr.2010.09.002} {\bibfield  {journal} {\bibinfo  {journal} {Journal of Magnetic Resonance}\ }\textbf {\bibinfo {volume} {207}},\ \bibinfo {pages} {251} (\bibinfo {year} {2010})}\BibitemShut {NoStop}%
\bibitem [{Note1()}]{Note1}%
  \BibitemOpen
  \bibinfo {note} {Calculations are bench-marked using a \protect \textit {MacBookPro18,2} equipped with a \protect \textit {M1~Max} processor with eight 3228~MHz \protect \textit {performance} cores, two 2064~MHz \protect \textit {efficiency} cores, and 32~GB of 512-bit LPDDR5 SDRAM memory.}\BibitemShut {Stop}%
\bibitem [{\citenamefont {Packer}\ \emph {et~al.}(2021)\citenamefont {Packer}, \citenamefont {Hobson}, \citenamefont {Holmes}, \citenamefont {Leggett}, \citenamefont {Glover}, \citenamefont {Brookes}, \citenamefont {Bowtell},\ and\ \citenamefont {Fromhold}}]{PhysRevApplied.15.064006}%
  \BibitemOpen
  \bibfield  {author} {\bibinfo {author} {\bibfnamefont {M.}~\bibnamefont {Packer}}, \bibinfo {author} {\bibfnamefont {P.}~\bibnamefont {Hobson}}, \bibinfo {author} {\bibfnamefont {N.}~\bibnamefont {Holmes}}, \bibinfo {author} {\bibfnamefont {J.}~\bibnamefont {Leggett}}, \bibinfo {author} {\bibfnamefont {P.}~\bibnamefont {Glover}}, \bibinfo {author} {\bibfnamefont {M.}~\bibnamefont {Brookes}}, \bibinfo {author} {\bibfnamefont {R.}~\bibnamefont {Bowtell}},\ and\ \bibinfo {author} {\bibfnamefont {T.}~\bibnamefont {Fromhold}},\ }\href {https://doi.org/10.1103/PhysRevApplied.15.064006} {\bibfield  {journal} {\bibinfo  {journal} {Physical Review Applied}\ }\textbf {\bibinfo {volume} {15}},\ \bibinfo {pages} {064006} (\bibinfo {year} {2021})}\BibitemShut {NoStop}%
\bibitem [{Note2()}]{Note2}%
  \BibitemOpen
  \bibinfo {note} {Another root exists at $P_{4,0}\left (x=0.362\right ){\approx }0$, however, this ring magnet design generates a much smaller region with high uniformity than the more separated design.}\BibitemShut {Stop}%
\bibitem [{\citenamefont {Bidinosti}\ \emph {et~al.}(2005)\citenamefont {Bidinosti}, \citenamefont {Kravchuk},\ and\ \citenamefont {Hayden}}]{10.1016/j.jmr.2005.07.003}%
  \BibitemOpen
  \bibfield  {author} {\bibinfo {author} {\bibfnamefont {C.}~\bibnamefont {Bidinosti}}, \bibinfo {author} {\bibfnamefont {I.}~\bibnamefont {Kravchuk}},\ and\ \bibinfo {author} {\bibfnamefont {M.}~\bibnamefont {Hayden}},\ }\href {https://doi.org/10.1016/j.jmr.2005.07.003} {\bibfield  {journal} {\bibinfo  {journal} {Journal of magnetic resonance (San Diego, Calif. : 1997)}\ }\textbf {\bibinfo {volume} {177}},\ \bibinfo {pages} {31} (\bibinfo {year} {2005})}\BibitemShut {NoStop}%
\bibitem [{\citenamefont {Limes}\ \emph {et~al.}(2020)\citenamefont {Limes}, \citenamefont {Foley}, \citenamefont {Kornack}, \citenamefont {Caliga}, \citenamefont {McBride}, \citenamefont {Braun}, \citenamefont {Lee}, \citenamefont {Lucivero},\ and\ \citenamefont {Romalis}}]{PhysRevApplied.14.011002}%
  \BibitemOpen
  \bibfield  {author} {\bibinfo {author} {\bibfnamefont {M.}~\bibnamefont {Limes}}, \bibinfo {author} {\bibfnamefont {E.}~\bibnamefont {Foley}}, \bibinfo {author} {\bibfnamefont {T.}~\bibnamefont {Kornack}}, \bibinfo {author} {\bibfnamefont {S.}~\bibnamefont {Caliga}}, \bibinfo {author} {\bibfnamefont {S.}~\bibnamefont {McBride}}, \bibinfo {author} {\bibfnamefont {A.}~\bibnamefont {Braun}}, \bibinfo {author} {\bibfnamefont {W.}~\bibnamefont {Lee}}, \bibinfo {author} {\bibfnamefont {V.}~\bibnamefont {Lucivero}},\ and\ \bibinfo {author} {\bibfnamefont {M.}~\bibnamefont {Romalis}},\ }\href {https://doi.org/10.1103/PhysRevApplied.14.011002} {\bibfield  {journal} {\bibinfo  {journal} {Physical Review Applied}\ }\textbf {\bibinfo {volume} {14}},\ \bibinfo {pages} {011002} (\bibinfo {year} {2020})}\BibitemShut {NoStop}%
\bibitem [{\citenamefont {Li}\ \emph {et~al.}(2017)\citenamefont {Li}, \citenamefont {Post}, \citenamefont {Kunc}, \citenamefont {Elliott},\ and\ \citenamefont {Paranthaman}}]{10.1016/j.scriptamat.2016.12.035}%
  \BibitemOpen
  \bibfield  {author} {\bibinfo {author} {\bibfnamefont {L.}~\bibnamefont {Li}}, \bibinfo {author} {\bibfnamefont {B.}~\bibnamefont {Post}}, \bibinfo {author} {\bibfnamefont {V.}~\bibnamefont {Kunc}}, \bibinfo {author} {\bibfnamefont {A.}~\bibnamefont {Elliott}},\ and\ \bibinfo {author} {\bibfnamefont {P.}~\bibnamefont {Paranthaman}},\ }\href {https://doi.org/10.1016/j.scriptamat.2016.12.035} {\bibfield  {journal} {\bibinfo  {journal} {Scripta Materialia}\ }\textbf {\bibinfo {volume} {135}} (\bibinfo {year} {2017})}\BibitemShut {NoStop}%
\bibitem [{\citenamefont {Tiismus}\ \emph {et~al.}(2022)\citenamefont {Tiismus}, \citenamefont {Kallaste}, \citenamefont {Vaimann},\ and\ \citenamefont {Rassõlkin}}]{TIISMUS2022102778}%
  \BibitemOpen
  \bibfield  {author} {\bibinfo {author} {\bibfnamefont {H.}~\bibnamefont {Tiismus}}, \bibinfo {author} {\bibfnamefont {A.}~\bibnamefont {Kallaste}}, \bibinfo {author} {\bibfnamefont {T.}~\bibnamefont {Vaimann}},\ and\ \bibinfo {author} {\bibfnamefont {A.}~\bibnamefont {Rassõlkin}},\ }\href {https://doi.org/https://doi.org/10.1016/j.addma.2022.102778} {\bibfield  {journal} {\bibinfo  {journal} {Additive Manufacturing}\ }\textbf {\bibinfo {volume} {55}},\ \bibinfo {pages} {102778} (\bibinfo {year} {2022})}\BibitemShut {NoStop}%
\bibitem [{\citenamefont {Packer}\ \emph {et~al.}(2020)\citenamefont {Packer}, \citenamefont {Hobson}, \citenamefont {Holmes}, \citenamefont {Leggett}, \citenamefont {Glover}, \citenamefont {Brookes}, \citenamefont {Bowtell},\ and\ \citenamefont {Fromhold}}]{PhysRevApplied.15.054004}%
  \BibitemOpen
  \bibfield  {author} {\bibinfo {author} {\bibfnamefont {M.}~\bibnamefont {Packer}}, \bibinfo {author} {\bibfnamefont {P.}~\bibnamefont {Hobson}}, \bibinfo {author} {\bibfnamefont {N.}~\bibnamefont {Holmes}}, \bibinfo {author} {\bibfnamefont {J.}~\bibnamefont {Leggett}}, \bibinfo {author} {\bibfnamefont {P.}~\bibnamefont {Glover}}, \bibinfo {author} {\bibfnamefont {M.}~\bibnamefont {Brookes}}, \bibinfo {author} {\bibfnamefont {R.}~\bibnamefont {Bowtell}},\ and\ \bibinfo {author} {\bibfnamefont {T.}~\bibnamefont {Fromhold}},\ }\href {https://doi.org/10.1103/PhysRevApplied.14.054004} {\bibfield  {journal} {\bibinfo  {journal} {Physical Review Applied}\ }\textbf {\bibinfo {volume} {14}},\ \bibinfo {pages} {054004} (\bibinfo {year} {2020})}\BibitemShut {NoStop}%
\bibitem [{\citenamefont {Hobson}\ \emph {et~al.}(2023)\citenamefont {Hobson}, \citenamefont {Holmes}, \citenamefont {Patel}, \citenamefont {Styles}, \citenamefont {Chalmers}, \citenamefont {Morley}, \citenamefont {Davis}, \citenamefont {Packer}, \citenamefont {Smith}, \citenamefont {Raudonyte}, \citenamefont {Holmes}, \citenamefont {Harrison}, \citenamefont {Woolger}, \citenamefont {Sims}, \citenamefont {Brookes}, \citenamefont {Bowtell},\ and\ \citenamefont {Fromhold}}]{10177829}%
  \BibitemOpen
  \bibfield  {author} {\bibinfo {author} {\bibfnamefont {P.~J.}\ \bibnamefont {Hobson}}, \bibinfo {author} {\bibfnamefont {N.}~\bibnamefont {Holmes}}, \bibinfo {author} {\bibfnamefont {P.}~\bibnamefont {Patel}}, \bibinfo {author} {\bibfnamefont {B.}~\bibnamefont {Styles}}, \bibinfo {author} {\bibfnamefont {J.}~\bibnamefont {Chalmers}}, \bibinfo {author} {\bibfnamefont {C.}~\bibnamefont {Morley}}, \bibinfo {author} {\bibfnamefont {A.}~\bibnamefont {Davis}}, \bibinfo {author} {\bibfnamefont {M.}~\bibnamefont {Packer}}, \bibinfo {author} {\bibfnamefont {T.~X.}\ \bibnamefont {Smith}}, \bibinfo {author} {\bibfnamefont {S.}~\bibnamefont {Raudonyte}}, \bibinfo {author} {\bibfnamefont {D.}~\bibnamefont {Holmes}}, \bibinfo {author} {\bibfnamefont {R.}~\bibnamefont {Harrison}}, \bibinfo {author} {\bibfnamefont {D.}~\bibnamefont {Woolger}}, \bibinfo {author} {\bibfnamefont {D.}~\bibnamefont {Sims}}, \bibinfo {author} {\bibfnamefont {M.~J.}\ \bibnamefont {Brookes}}, \bibinfo {author} {\bibfnamefont {R.}~\bibnamefont
  {Bowtell}},\ and\ \bibinfo {author} {\bibfnamefont {M.}~\bibnamefont {Fromhold}},\ }\href {https://doi.org/10.1109/TIM.2023.3293540} {\bibfield  {journal} {\bibinfo  {journal} {IEEE Transactions on Instrumentation and Measurement}\ }\textbf {\bibinfo {volume} {72}},\ \bibinfo {pages} {1} (\bibinfo {year} {2023})}\BibitemShut {NoStop}%
\bibitem [{\citenamefont {Holmes}\ \emph {et~al.}(2022)\citenamefont {Holmes}, \citenamefont {Rea}, \citenamefont {Chalmers}, \citenamefont {Leggett}, \citenamefont {Edwards}, \citenamefont {Nell}, \citenamefont {Pink}, \citenamefont {Patel}, \citenamefont {Wood}, \citenamefont {Murby}, \citenamefont {Woolger}, \citenamefont {Dawson}, \citenamefont {Mariani}, \citenamefont {Tierney}, \citenamefont {Mellor}, \citenamefont {O’Neill}, \citenamefont {Boto}, \citenamefont {Hill}, \citenamefont {Shah},\ and\ \citenamefont {Bowtell}}]{10.1038/s41598-022-17346-1}%
  \BibitemOpen
  \bibfield  {author} {\bibinfo {author} {\bibfnamefont {N.}~\bibnamefont {Holmes}}, \bibinfo {author} {\bibfnamefont {M.}~\bibnamefont {Rea}}, \bibinfo {author} {\bibfnamefont {J.}~\bibnamefont {Chalmers}}, \bibinfo {author} {\bibfnamefont {J.}~\bibnamefont {Leggett}}, \bibinfo {author} {\bibfnamefont {L.}~\bibnamefont {Edwards}}, \bibinfo {author} {\bibfnamefont {P.}~\bibnamefont {Nell}}, \bibinfo {author} {\bibfnamefont {S.}~\bibnamefont {Pink}}, \bibinfo {author} {\bibfnamefont {P.}~\bibnamefont {Patel}}, \bibinfo {author} {\bibfnamefont {J.}~\bibnamefont {Wood}}, \bibinfo {author} {\bibfnamefont {N.}~\bibnamefont {Murby}}, \bibinfo {author} {\bibfnamefont {D.}~\bibnamefont {Woolger}}, \bibinfo {author} {\bibfnamefont {E.}~\bibnamefont {Dawson}}, \bibinfo {author} {\bibfnamefont {C.}~\bibnamefont {Mariani}}, \bibinfo {author} {\bibfnamefont {T.}~\bibnamefont {Tierney}}, \bibinfo {author} {\bibfnamefont {S.}~\bibnamefont {Mellor}}, \bibinfo {author} {\bibfnamefont {G.}~\bibnamefont {O’Neill}}, \bibinfo
  {author} {\bibfnamefont {E.}~\bibnamefont {Boto}}, \bibinfo {author} {\bibfnamefont {R.}~\bibnamefont {Hill}}, \bibinfo {author} {\bibfnamefont {V.}~\bibnamefont {Shah}},\ and\ \bibinfo {author} {\bibfnamefont {R.}~\bibnamefont {Bowtell}},\ }\href {https://doi.org/10.1038/s41598-022-17346-1} {\bibfield  {journal} {\bibinfo  {journal} {Scientific Reports}\ }\textbf {\bibinfo {volume} {12}} (\bibinfo {year} {2022})}\BibitemShut {NoStop}%
\bibitem [{\citenamefont {Jain}\ \emph {et~al.}(2024)\citenamefont {Jain}, \citenamefont {Sägesser}, \citenamefont {Hrmo}, \citenamefont {Torkzaban}, \citenamefont {Stadler}, \citenamefont {Oswald}, \citenamefont {Axline}, \citenamefont {Bautista-Salvador}, \citenamefont {Ospelkaus}, \citenamefont {Kienzler},\ and\ \citenamefont {Home}}]{Jain_2024}%
  \BibitemOpen
  \bibfield  {author} {\bibinfo {author} {\bibfnamefont {S.}~\bibnamefont {Jain}}, \bibinfo {author} {\bibfnamefont {T.}~\bibnamefont {Sägesser}}, \bibinfo {author} {\bibfnamefont {P.}~\bibnamefont {Hrmo}}, \bibinfo {author} {\bibfnamefont {C.}~\bibnamefont {Torkzaban}}, \bibinfo {author} {\bibfnamefont {M.}~\bibnamefont {Stadler}}, \bibinfo {author} {\bibfnamefont {R.}~\bibnamefont {Oswald}}, \bibinfo {author} {\bibfnamefont {C.}~\bibnamefont {Axline}}, \bibinfo {author} {\bibfnamefont {A.}~\bibnamefont {Bautista-Salvador}}, \bibinfo {author} {\bibfnamefont {C.}~\bibnamefont {Ospelkaus}}, \bibinfo {author} {\bibfnamefont {D.}~\bibnamefont {Kienzler}},\ and\ \bibinfo {author} {\bibfnamefont {J.}~\bibnamefont {Home}},\ }\href {https://doi.org/10.1038/s41586-024-07111-x} {\bibfield  {journal} {\bibinfo  {journal} {Nature}\ }\textbf {\bibinfo {volume} {627}},\ \bibinfo {pages} {510–514} (\bibinfo {year} {2024})}\BibitemShut {NoStop}%
\end{thebibliography}%

\end{document}